\documentclass[twocolumn,superscriptaddress,floatfix,aps,prd,showpacs]{revtex4}

\usepackage{graphicx}

\begin{document}

\title{Relativistic black hole-neutron star binaries in
quasiequilibrium: effects of the black hole excision boundary
condition}

\author{Keisuke Taniguchi}
\affiliation{Department of Physics, University of Illinois at
  Urbana-Champaign, Urbana, Illinois 61801, USA}
\author{Thomas W. Baumgarte}
\altaffiliation{Also at: Department of Physics, University of Illinois
  at Urbana-Champaign, Urbana, Illinois 61801, USA}
\affiliation{Department of Physics and Astronomy, Bowdoin College,
  Brunswick, Maine 04011, USA}
\author{Joshua A. Faber}
\affiliation{Department of Physics, University of Illinois at
  Urbana-Champaign, Urbana, Illinois 61801, USA}
\author{Stuart L. Shapiro}
\altaffiliation{Also at: Department of Astronomy and NCSA, University
  of Illinois at Urbana-Champaign, Urbana, Illinois 61801, USA}
\affiliation{Department of Physics, University of Illinois at
  Urbana-Champaign, Urbana, Illinois 61801, USA}

\date{\today}

\begin{abstract}
We construct new models of black hole-neutron star binaries in
quasiequilibrium circular orbits by solving Einstein's constraint
equations in the conformal thin-sandwich decomposition together with
the relativistic equations of hydrostationary equilibrium. We adopt
maximal slicing, assume spatial conformal flatness, and impose
equilibrium boundary conditions on an excision surface (i.e., the
apparent horizon) to model the black hole. In our previous treatment
we adopted a ``leading-order" approximation for a parameter related to
the black-hole spin in these boundary conditions to construct
approximately nonspinning black holes. Here we improve on the models
by computing the black hole's quasilocal spin angular momentum and
setting it to zero. As before, we adopt a polytropic equation of state
with adiabatic index $\Gamma=2$ and assume the neutron star to be
irrotational. In addition to recomputing several sequences for
comparison with our earlier results, we study a wider range of neutron
star masses and binary mass ratios. To locate the innermost stable
circular orbit we search for turning points along both the binding
energy and total angular momentum curves for these sequences. Unlike
for our previous approximate boundary condition, these two minima now
coincide. We also identify the formation of cusps on the neutron star
surface, indicating the onset of tidal disruption. Comparing these two
critical binary separations for different mass ratios and neutron star
compactions we distinguish those regions that will lead to a tidal
disruption of the neutron star from those that will result in the
plunge into the black hole of a neutron star more or less intact,
albeit distorted by tidal forces.
\end{abstract}


\maketitle

\section{Introduction}

Coalescing black hole-neutron star (BHNS) binaries, as well as other
compact binaries composed of neutron stars and/or black holes, are
among the most promising sources of gravitational waves for both
ground-based \cite{LIGO,GEO,TAMA,VIRGO} and space-based laser
interferometers \cite{LISA,DECIGO}. BHNS binary mergers are also
candidate central engines of short-hard gamma-ray bursts (SGRBs) (see,
e.g. \cite{LeeR07} and references cited therein). The remnants of both
BHNS binary mergers \cite{FaberBST06,ShibaU0607} and binary neutron
star mergers \cite{ShibaT06,PriceR06,OechsJ06,HMNS} are feasible
progenitors for SGRBs because both may result in black holes
surrounded by hot, massive accretion disks with very little, if any,
baryon contamination along the polar symmetry axis.

Motivated by these factors, considerable effort has gone into the
study of BHNS binaries. Most approaches to date assume Newtonian
gravity in either some or all aspects of the calculation (see,
e.g.~\cite{Chand69,Fishb73,LaiRS93,LaiW96,TanigN96,Shiba96,UryuE99,WiggiL00,IshiiSM05,Mille05}
for quasiequilibrium calculations and
\cite{Mashh75,CarteL83,Marck83,LeeK99,Lee00,RosswSW04,KobayLPM04,RantsKLR07}
for dynamical simulations). More recently, several groups have also
studied BHNS binaries in a fully relativistic framework, both for
quasiequilibrium models
\cite{Mille01,BaumgSS04,TanigBFS05,TanigBFS06,Grand06,TanigBFS07}
and dynamical simulations
\cite{FaberBSTR06,FaberBST06,SopueSL06,LofflRA06,ShibaU0607}.

Our group has pursued a systematic approach to developing increasingly
realistic models of BHNS binaries in quasiequilibrium circular
orbits. Our first studies
\cite{BaumgSS04,TanigBFS05,FaberBSTR06,FaberBST06} assumed extreme
mass ratios, i.e., black hole masses that are much greater than the
neutron star mass. While this is a very natural first step from a
computational point of view, binaries with comparable masses are much
more interesting from the perspective of ground-based gravitational
wave observations and for the launching of SGRBs. More recently we
have therefore relaxed this assumption and have extended our results
to the case of comparable-mass BHNS binaries
\cite{TanigBFS06,TanigBFS07}.

Specifically, in \cite{TanigBFS07} (hereafter Paper I) we constructed
quasiequilibrium models by solving Einstein's constraint equations in
the conformal thin-sandwich formalism, assuming conformal flatness and
maximal slicing, together with the relativistic equations of
hydrostationary equilibrium. We accounted for the black hole by
excising a coordinate sphere and imposing the equilibrium black-hole
boundary conditions of Cook and Pfeiffer \cite{CookP04}. This original
version implemented a ``leading-order" approximation to nonspinning
black holes, which equates an otherwise undetermined spin parameter
$\Omega_r$ that appears in the boundary condition for the shift vector
with the orbital angular velocity seen by an inertial observer at
infinity, $\Omega$. As for the original irrotational binary black hole
models of \cite{CookP04}, this condition does not lead to simultaneous
turning points of the binding energy and the total angular momentum in
constant-mass sequences in Paper I. Such simultaneous turning points
are expected for those sequences if they are truly in
quasiequilibrium \cite{dMOmegadJ}. An improvement over this condition,
namely to iterate over $\Omega_r$ until the quasilocal spin angular
momentum of the black hole vanishes, was suggested and implemented for
binary black holes by \cite{CaudiCGP06}.

In this paper we reconstruct quasiequilibrium models of BHNS binaries
using the same techniques as in Paper I, but with the improved black
hole spin angular velocity condition as suggested by
\cite{CaudiCGP06}. We then compute sequences of BHNS binaries in
quasicircular orbits for a wider range of neutron star masses and
binary mass ratios than in Paper I, focusing our attention on
irrotational neutron stars orbiting nonspinning black holes. Here we
focus only on the irrotational state for the neutron star because it
is astrophysically considered to be more realistic in a BHNS binary
\cite{Kocha92,BildsC92,FaberBSTR06}. On the other hand, we will
compute the case of spinning black holes in future work. As was the
case for the irrotational black hole binaries constructed in
\cite{CaudiCGP06}, we find that this improved condition for the spin
parameter of the black hole $\Omega_r$ does lead to simultaneous
turning points in the binding energy and the total angular momentum
along constant-mass sequences.

The paper is organized as follows. We briefly review the basic
equations in Section II. We present numerical results in Section III,
and outline some qualitative considerations concerning the fate of
BHNS binaries in Section IV. In Section V we summarize our
findings. Throughout this paper we adopt geometrized units with
$G=c=1$, where $G$ denotes the gravitational constant and $c$ the
speed of light. Latin and Greek indices denote purely spatial and
spacetime components, respectively.

\section{Formulation}

In this Section we review the equations we solve to construct a
quasiequilibrium BHNS binary. The equations are very similar to those
in Paper I, but, as we explain below, we have improved both our
algorithm for the solution of the gravitational field equations and
the method to compute the spin parameter $\Omega_r$ that appears in
the black hole boundary condition for the shift vector. For a more
detailed discussion about the formalism we use here, we refer to Paper
I, the review articles \cite{Cook00,BaumgS03}, and Sec.~II of
\cite{GourgGTMB01} for the hydrostatics.

\subsection{Gravitational field equations}
\label{Sec:IIA}

The line element in $3+1$ form is written as
\begin{eqnarray}
  ds^2 &=& g_{\mu \nu} dx^{\mu} dx^{\nu} \nonumber \\
  &=& -\alpha^2 dt^2 + \gamma_{ij} (dx^i +\beta^i dt)
  (dx^j +\beta^j dt),
\end{eqnarray}
where $g_{\mu \nu}$ is the spacetime metric, $\alpha$ the lapse
function, $\beta^i$ the shift vector, and $\gamma_{ij}$ the spatial
metric induced on a spatial slice $\Sigma$. The spatial metric
$\gamma_{ij}$ is further decomposed according to $\gamma_{ij} = \psi^4
\tilde{\gamma}_{ij}$, where $\psi$ denotes the conformal factor and
$\tilde{\gamma}_{ij}$ the conformal background spatial metric, defined
such that $\det \tilde{\gamma}=1$ in Cartesian coordinates. We also
decompose the extrinsic curvature $K^{ij}$ into a trace $K$ and a
traceless part $\tilde{A}^{ij}$ according to
\begin{equation}
  K^{ij} = \psi^{-10} \tilde{A}^{ij} +\frac{1}{3} \gamma^{ij} K.
\end{equation}
The Hamiltonian constraint then becomes
\begin{equation} \label{eq:ham}
  \tilde{\nabla}^2 \psi = -2\pi \psi^5 \rho +\frac{1}{8} \psi
  \tilde{R} +\frac{1}{12} \psi^5 K^2 - \frac{1}{8} \psi^{-7}
  \tilde{A}_{ij} \tilde{A}^{ij}. 
\end{equation}
Here $\tilde{\nabla}^2 = \tilde \gamma^{ij} \tilde \nabla_i \tilde
\nabla_j$ is the covariant Laplace operator, $\tilde \nabla_i$ the
covariant derivative, $\tilde{R}_{ij}$ the Ricci tensor, and
$\tilde{R}=\tilde{\gamma}^{ij} \tilde{R}_{ij}$ the scalar curvature,
all associated with the conformal background metric
$\tilde{\gamma}_{ij}$.

We employ the conformal thin-sandwich decomposition of the Einstein
equations \cite{York99}. In this decomposition, we use the evolution
equation for the spatial metric to express the traceless part of the
extrinsic curvature in terms of the time derivative of the background
metric, $\tilde{u}_{ij}\equiv\partial_t \tilde{\gamma}_{ij}$, and the
gradients of the shift vector. Under the assumption of equilibrium,
i.e., $\tilde{u}_{ij}=0$ in a corotating coordinate system, the
traceless part of the extrinsic curvature reduces to
\begin{equation}
  \tilde{A}^{ij} =\frac{\psi^6}{2\alpha} \Bigl( \tilde{\nabla}^i
  \beta^j +\tilde{\nabla}^j \beta^i -\frac{2}{3} \tilde{\gamma}^{ij}
  \tilde{\nabla}_k \beta^k \Bigr). \label{eq:taij}
\end{equation}
Inserting Eq.~(\ref{eq:taij}) into the momentum constraint we
obtain
\begin{eqnarray} \label{eq:mom}
  &&\tilde{\nabla}^2 \beta^i +\frac{1}{3} \tilde{\nabla}^i
  (\tilde{\nabla}_j \beta^j) +\tilde{R}^i_j \beta^j \nonumber \\
  &&= 16\pi \alpha \psi^4 j^i
  + 2\tilde{A}^{ij} \tilde{\nabla}_j (\alpha \psi^{-6})
  +\frac{4}{3} \alpha \tilde{\gamma}^{ij} \tilde{\nabla}_j K.
\end{eqnarray}
For the construction of quasiequilibrium data it is also reasonable to
assume $\partial_t K =0$ in a corotating coordinate system. The trace
of the evolution equation for the extrinsic curvature then yields
\begin{eqnarray} \label{eq:Kdot}
  \tilde{\nabla}^2 \alpha &=& 4\pi \alpha \psi^4 (\rho +S)
  +\frac{1}{3} \alpha \psi^4 K^2 + \psi^4 \beta^i \tilde{\nabla}_i K
  \nonumber \\
  &&+ \alpha \psi^{-8} \tilde{A}_{ij} \tilde{A}^{ij}
  - 2 \tilde{\gamma}^{ij} \tilde{\nabla}_i \alpha \tilde{\nabla}_j
  \ln \psi.
\end{eqnarray}

The matter terms on the right-hand side of Eqs.~(\ref{eq:ham}),
(\ref{eq:mom}), and (\ref{eq:Kdot}) are derived from the
projections of the stress-energy tensor $T_{\mu \nu}$ into the spatial
slice $\Sigma$. Denoting the future-oriented unit normal to $\Sigma$
as $n_{\mu}$, the relevant projections of $T_{\mu \nu}$ are
\begin{eqnarray}
  &&\rho = n_{\mu} n_{\nu} T^{\mu \nu}, \\
  &&j^i = -\gamma^i_{\mu} n_{\nu} T^{\mu \nu}, \\
  &&S_{ij} = \gamma_{i \mu} \gamma_{j \nu} T^{\mu \nu}, \\
  &&S = \gamma^{ij} S_{ij}.
\end{eqnarray}
Here we write the stress-energy tensor as
\begin{equation}
  T_{\mu \nu} = (\rho_0 +\rho_i +P) u_{\mu} u_{\nu} + P g_{\mu \nu},
\end{equation}
assuming an ideal fluid. The quantity $u_{\mu}$ is the fluid
4-velocity, $\rho_0$ the baryon rest-mass density, $\rho_i$ the
internal energy density, and $P$ the pressure.

Equations (\ref{eq:ham}), (\ref{eq:mom}) and (\ref{eq:Kdot}) provide
equations for the lapse function $\alpha$, the shift vector $\beta^i$,
and the conformal factor $\psi$, while $\tilde A^{ij}$ can be found
from Eq.~(\ref{eq:taij}). The conformally related spatial metric
$\tilde{\gamma}_{ij}$ and the trace of the extrinsic curvature $K$
remain freely specifiable, and have to be chosen before we can solve
the above equations (note that we have already set the time
derivatives of these quantities, which are also freely specifiable, to
zero). As in Paper I we assume a flat background
$\tilde{\gamma}_{ij} = \eta_{ij}$, where $\eta_{ij}$ denotes a flat
spatial metric, and maximal slicing, $K=0$. In Cartesian coordinates,
Eqs.~(\ref{eq:ham}), (\ref{eq:mom}) and (\ref{eq:Kdot}) can then be
written as
\begin{eqnarray}
  &&\underline{\Delta} \psi = -2\pi \psi^5 \rho - \frac{1}{8} \psi^{-7}
  \tilde{A}_{ij} \tilde{A}^{ij}, \label{eq:ham_constr} \\
  &&\underline{\Delta} \beta^i +\frac{1}{3} \partial^i
  (\partial_j \beta^j) = 16\pi \Phi \psi^3 j^i
  + 2\tilde{A}^{ij} \partial_j (\Phi \psi^{-7}),
  \label{eq:mom_constr} \\
  &&\underline{\Delta} \Phi = 2\pi \Phi \psi^4 (\rho +2S)
  + \frac{7}{8} \Phi \psi^{-8} \tilde{A}_{ij} \tilde{A}^{ij},
  \label{eq:trace_evolv}
\end{eqnarray}
where $\underline{\Delta}$ and $\partial_i$ denote the flat Laplace
operator and the partial derivative, and
$\Phi \equiv \alpha \psi$. Eq.~(\ref{eq:taij}) becomes
\begin{equation}
  \tilde{A}^{ij} =\frac{\psi^7}{2\Phi} \Bigl( \partial^i
  \beta^j +\partial^j \beta^i -\frac{2}{3} \eta^{ij}
  \partial_k \beta^k \Bigr).
\end{equation}
For numerical purposes we further decompose the variables and their
equations into parts associated with the black hole and the neutron
star. For details of this decomposition we refer to Appendix A
of Paper I \cite{TanigBFS07}.

In Paper I we solved directly for the lapse function $\alpha$, the
conformal factor $\psi$ and the shift vector $\beta^i$. Instead of
solving Eq.~(\ref{eq:Kdot}) for the lapse, we now solve
Eq.~(\ref{eq:trace_evolv}) for the combination $\Phi = \alpha
\psi$. This choice is quite common (see e.g.~\cite{Cook00,BaumgS03}),
and has the advantage of eliminating the source term $-2\eta^{ij}
\partial_i \alpha \partial_j \ln \psi$ on the right-hand side of
Eq.~(\ref{eq:Kdot}) above, or equivalently Eq. (9) in Paper I. This
term falls off like $1/r^4$, and hence more slowly than $\tilde{A}_{ij}
\tilde{A}^{ij}$. Eliminating this term therefore enhances the accuracy
of the numerical solution. We will quantify the improvement in
Section \ref{Sec:IIIB}.

\subsection{Boundary conditions}

In order to solve the gravitational field equations
(\ref{eq:ham_constr}), (\ref{eq:mom_constr}), and
(\ref{eq:trace_evolv}), we have to impose appropriate boundary
conditions on two different boundaries: outer boundaries at spatial
infinity and inner boundaries on the black hole horizons.

The boundary conditions at spatial infinity follow from the assumption
of asymptotic flatness.  With the help of a radial coordinate
transformation $u=1/r$ in the external computational domain, our
computational grid extends to spatial infinity
\cite{BonazGM98,GourgGTMB01}, and we can impose the exact boundary
conditions
\begin{eqnarray}
  &&\psi |_{r \rightarrow \infty} = 1, \\
  &&\beta^i |_{r \rightarrow \infty} = (\mbox{\boldmath $\Omega$}
  \times \mbox{\boldmath $R$})^i, \\
  &&\Phi |_{r \rightarrow \infty} = 1.
\end{eqnarray}
Here $\Omega$ is the orbital angular velocity of the binary system
measured at infinity, and $\mbox{\boldmath $R$}=(X,~Y,~Z)$ is a
Cartesian coordinate centered on the center of mass of the binary
system.  We express the shift vector $\beta^i$ in a corotating
coordinate system that we adopt throughout our calculation. In an
inertial coordinate system, the shift vector would tend to zero at
spatial infinity, while in the corotating coordinate system of the
numerical code the shift vector diverges at spatial infinity. For
computational purposes, it is therefore convenient to write the shift
vector as a sum of the rotational shift term $\beta^i_{\rm rot} \equiv
(\mbox{\boldmath $\Omega$} \times \mbox{\boldmath $R$})^i$ and a
residual part (which tends to zero at spatial infinity), and solve the
equations only for the latter.

The inner boundary conditions arise from the excision of the black
hole interior. The assumption that the black hole is in equilibrium
leads to a set of boundary conditions for the conformal factor and
shift vector \cite{CookP04} (see also \cite{Cook02,CaudiCGP06} as well
as the related isolated horizon formalism,
e.g.~\cite{AshteK04,GourgJ06}). The boundary condition for the
conformal factor is
\begin{equation}
  \tilde{s}^k \tilde{\nabla}_k \ln \psi \Bigl|_{\cal S}
  =-{1 \over 4} (\tilde{h}^{ij} \tilde{\nabla}_i \tilde{s}_j
  -\psi^2 J) \Bigl|_{\cal S}, \label{eq:psi_bc}
\end{equation}
where $s^i \equiv \psi^{-2} \tilde{s}^i$ is the outward pointing unit
vector normal to the excision surface and $h_{ij}$ is the induced
metric on the excision surface,
$h_{ij} \equiv \psi^4 \tilde{h}_{ij} = \gamma_{ij} - s_i s_j$. The
quantity $J$ is computed from the projection of the extrinsic
curvature $K_{ij}$ as $J \equiv h^{ij} K_{ij}$. The boundary
condition on the normal component of the shift vector is
\begin{equation}
  \beta_{\perp} |_{\cal S} = \alpha |_{\cal S}. \label{eq:beta_perp}
\end{equation}
The tangential components must form a conformal Killing vector of the
conformal metric $\tilde h_{ij}$ on the excision surface (see
\cite{CookP04}). This can be achieved by choosing them to be Killing
vectors of a 2-sphere,
\begin{equation}
  \beta_{\parallel}^i |_{\cal S} = \epsilon^{i}_{jk} \Omega_r^j x^k.
  \label{eq:beta_para}
\end{equation}
Here $\Omega_r^j$ is a freely specifiable vector, related to the
black-hole spin, that we take to be aligned with the $Z$-axis, and
$x^k$ is a Cartesian coordinate centered on the 2-sphere.

We assume the excision surface to be a coordinate sphere. In our
previous treatment we implemented a ``leading-order" approximation to
nonspinning black holes and set $\Omega_r = \Omega$, where $\Omega$ is
the orbital angular speed (compare \cite{CookP04}). Following
\cite{CaudiCGP06} we now iterate over $\Omega_r$ until the black
hole's quasilocal spin angular momentum
\begin{equation}
  S = {1 \over 8\pi} \oint_{\cal S} (K_{ij} -\gamma_{ij} K)
  \xi^j d^2 S^i
\end{equation}
vanishes. Here $\xi^i$ is an approximate Killing vector of $h_{ij}$
that we find by solving the Killing transport equations as described
in \cite{CaudiCGP06} (see also \cite{DreyeKSS03} for more detailed
descriptions and \cite{CookW07} for alternative methods for finding
approximate Killing vectors of closed 2-surfaces).

According to \cite{CookP04}, the boundary condition on the lapse
function can be chosen freely. In this paper, we choose a Neumann
boundary condition
\begin{equation} \label{eq:lapse_bound}
  \tilde{s}^i \partial_i \Phi \Bigl|_{\cal S} =0
\end{equation}
on the excised surface.

We refer to Sections II.E and II.F of Paper I for a discussion of how
the orbital angular velocity, the center of rotation, and several
global quantities including total angular momentum and mass are
computed.

\subsection{Numerical Method}

As in Paper I \cite{TanigBFS07}, we construct our numerical code based
on the {\sc Lorene} spectral-methods library routines developed by the
Meudon relativity group \cite{Lorene}. In our code, the computational
space is broken into multiple domains. Each domain around the
black hole is covered by $N_r \times N_{\theta} \times N_{\phi}=41
\times 33 \times 32$ or $49 \times 37 \times 36$ collocation points,
while those around the neutron star are covered by $25 \times 17 \times
16$ collocation points. Here $N_r$, $N_{\theta}$, and $N_{\phi}$
denote the number of collocation points for the radial, polar, and
azimuthal directions, respectively. We use a larger number of
collocation points for the black hole domains than for the neutron
star domains because the source terms of the black hole equations are
sensitive to the resolution of the neutron star. Since the black hole
domains are centered on the black hole, we need a higher angular
resolution to adequately resolve these source terms. The neutron star
equations, on the other hand, have large source terms only near the
neutron star. Since the neutron star domains are centered on the
neutron star, a more modest angular resolution is sufficient to
resolve these terms.

We refer to Appendix A of Paper I for a detailed discussion of
the decomposition of the equations and their source terms.

\section{Numerical results}

\begin{figure}[ht]
\vspace{0.2cm}
\begin{center}
  \includegraphics[width=8cm]{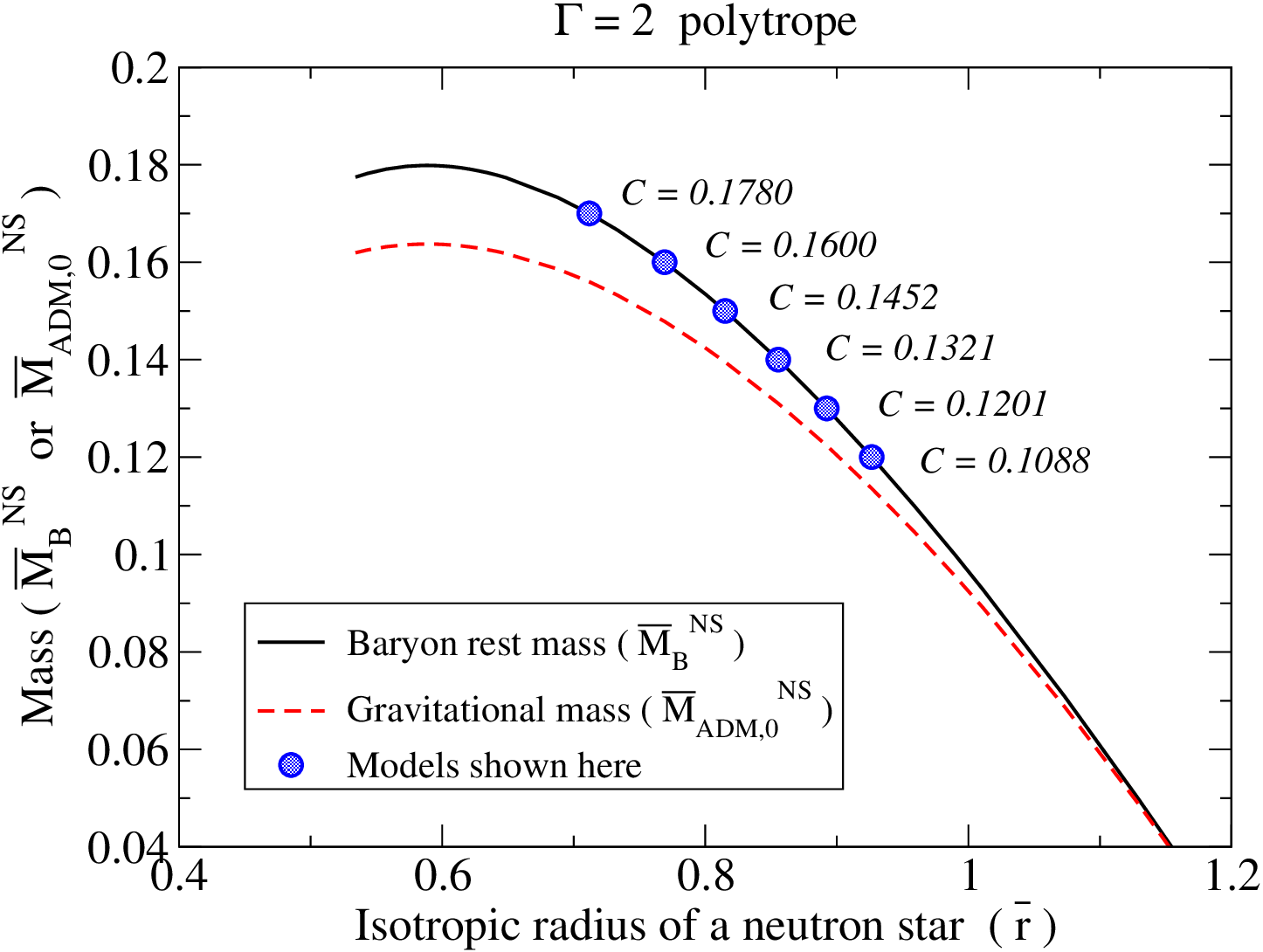}
\end{center}
\caption{Mass-radius relation for a spherical, polytropic star with
  adiabatic index $\Gamma=2$. The horizontal axis denotes the
  isotropic radius and the vertical axis the baryon rest mass (solid
  curve) and gravitational mass (dashed curve), where the radius and
  masses are in polytropic units. The gravitational mass is equivalent
  to the ADM mass for an isolated, spherical neutron star. Filled
  circles denote the models we choose in this paper. The compaction
  parameter ${\cal C}$ for each model is also shown.}
\label{fig:mass-radius}
\end{figure}

Throughout this paper, we model the neutron-star equation of state by
the polytropic relation $P=\kappa \rho_0^{\Gamma}$, where $P$ is the
pressure, $\rho_0$ the baryon rest-mass density, $\Gamma$ the
adiabatic index, and $\kappa$ a constant. We choose $\Gamma=2$ for the
adiabatic index, and compute several different constant-mass inspiral
sequences. Specifically the rest mass of the neutron star and
irreducible mass of the black hole are kept constant along each
sequence. We focus on baryon rest masses for neutron stars in the
range of $0.12 \le \bar{M}_{\rm B}^{\rm NS} \le 0.17$ (see
Fig.~\ref{fig:mass-radius}). Here and in the following we normalize
all dimensional quantities in terms of the polytropic length scale
$R_{\rm poly} \equiv \kappa^{1/(2\Gamma - 2)}$,
e.g.~$\bar M_{\rm B}^{\rm NS} = M_{\rm B}^{\rm NS}/R_{\rm poly}$. In
terms of compaction, our models are in the range
$0.1088 \le {\cal C} \le 0.1780$, where
\begin{equation}
  {\cal C} \equiv \frac{M_{\rm ADM,0}^{\rm NS}}{R_0}
\end{equation}
is the compaction of a spherical neutron star,
$M_{\rm ADM,0}^{\rm NS}$ the neutron star's ADM mass in isolation, and
$R_0$ its areal radius. Our most compact polytropic model with
$\bar{M}_{\rm B}^{\rm NS}=0.17$ is very close to the maximum baryon
rest mass for spherical $\Gamma=2$ polytropes in isolation,
$\bar{M}_{\rm B}^{\rm max} = 0.180$.

\vspace{0.2cm}

We consider mass ratios in the range $1 \le \hat{q} \le 10$,
where we define the mass ratio as
\begin{equation}
  \hat{q} \equiv \frac{M_{\rm irr}^{\rm BH}}{M_{\rm ADM,0}^{\rm NS}},
\end{equation}
i.e., the ratio of the irreducible mass of the black hole
($M_{\rm irr}^{\rm BH}$) to the ADM mass of a spherical, isolated
neutron star ($M_{\rm ADM,0}^{\rm NS}$). Here the irreducible mass
$M_{\rm irr}^{\rm BH}$ is identical to the ADM mass for an isolated
nonspinning black hole. Note again that we fix the irreducible mass of
the black hole and the baryon rest mass of the neutron star for the
construction of constant-mass sequences. For the definition of the
mass ratio, however, we use the ADM mass of a spherical isolated
neutron star $M_{\rm ADM,0}^{\rm NS}$, because this turns out to be
more convenient for comparisons with third-order post-Newtonian (3PN)
results \cite{Blanc02}.

We tabulate our numerical results in Appendix \ref{appendix:tables}.

\subsection{Configurations}

\begin{figure}[ht]
\vspace{0.2cm}
\begin{center}
  \includegraphics[width=8cm]{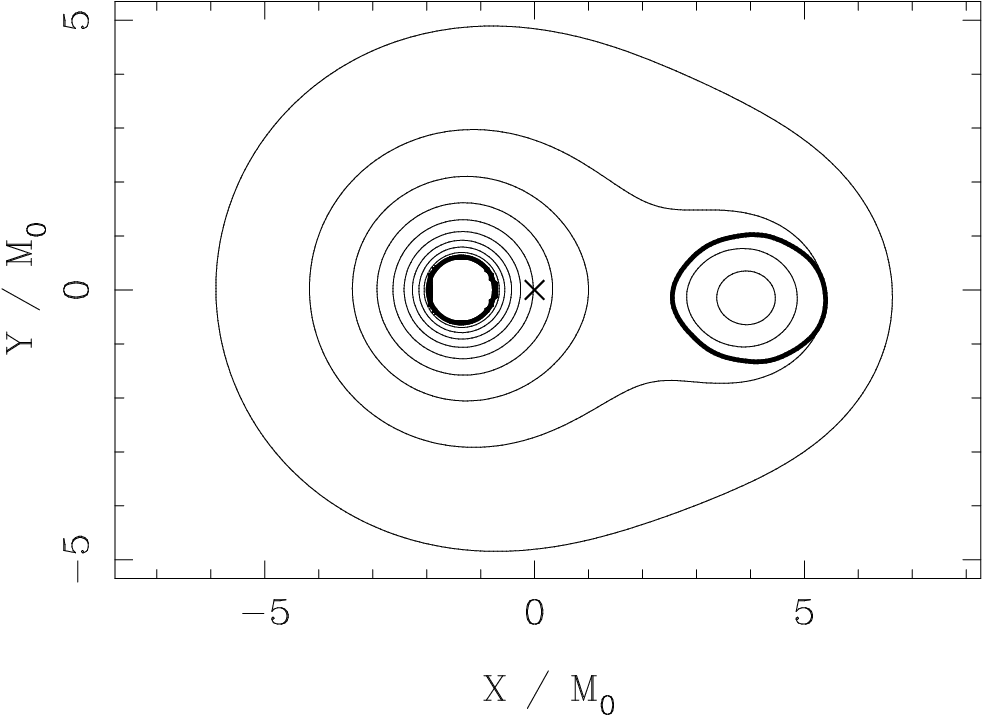}
\end{center}
\caption{Contours of the conformal factor $\psi$ in the equatorial
  plane for our innermost configuration with binary mass ratio
  $\hat{q}=3$ and neutron-star baryon rest mass $\bar{M}_{\rm B}^{\rm
  NS}=0.15$. The cross ``$\times$'' indicates the position of the
  rotation axis.}
\label{fig:confo}
\end{figure}

We show contours of the conformal factor $\psi$ for a BHNS binary
with mass ratio $\hat{q}=3$ and neutron-star baryon rest mass
$\bar{M}_{\rm B}^{\rm NS}=0.15$ in Fig. \ref{fig:confo}. This figure
represents the configuration at the smallest orbital separation for
which our code converged. The thick sold circle on the left-hand side
denotes the position of the excised surface (the apparent horizon),
while that on the right-hand side denotes the position of the
neutron-star surface. The value of $\psi$ on the excised surface is
not constant because a Neumann boundary condition (\ref{eq:psi_bc})
is applied.

\subsection{Binding energy and total angular momentum} \label{Sec:IIIB}

\begin{figure}[ht]
\vspace{0.2cm}
\begin{center}
  \includegraphics[width=8cm]{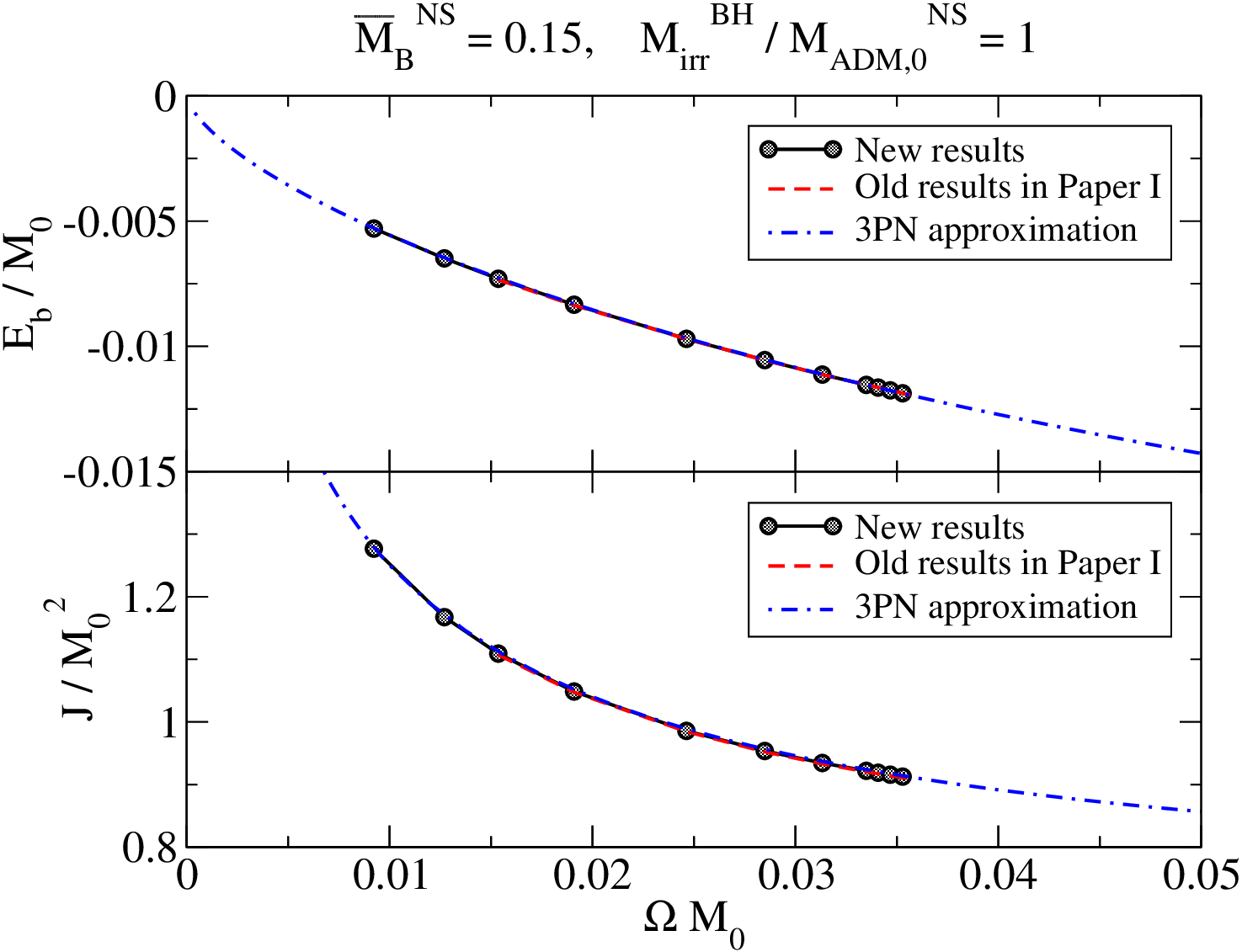}
\end{center}
\caption{Binding energy $E_{\rm b}/M_0$ (upper panel) and total
  angular momentum $J/M_0^2$ (lower panel) as a function of
  $\Omega M_0$ for binaries of mass ratio $\hat{q}=1$ and neutron-star
  mass $\bar{M}_{\rm B}^{\rm NS}=0.15$. The solid line with filled
  circles shows new results, and the dashed line the old results in
  Paper I. The dotted-dashed line denotes the results of the 3PN
  approximation \protect{\cite{Blanc02}}. The numerical sequences end
  due to cusp formation -- and hence the onset of tidal disruption --
  before the binary reaches the innermost stable circular orbit at the
  turning point of the binding energy. 3PN sequences all exhibit
  turning points and cannot reveal cusps.}
\label{fig:ej-o1M015}
\end{figure}

\begin{figure}[ht]
\vspace{0.2cm}
\begin{center}
  \includegraphics[width=8cm]{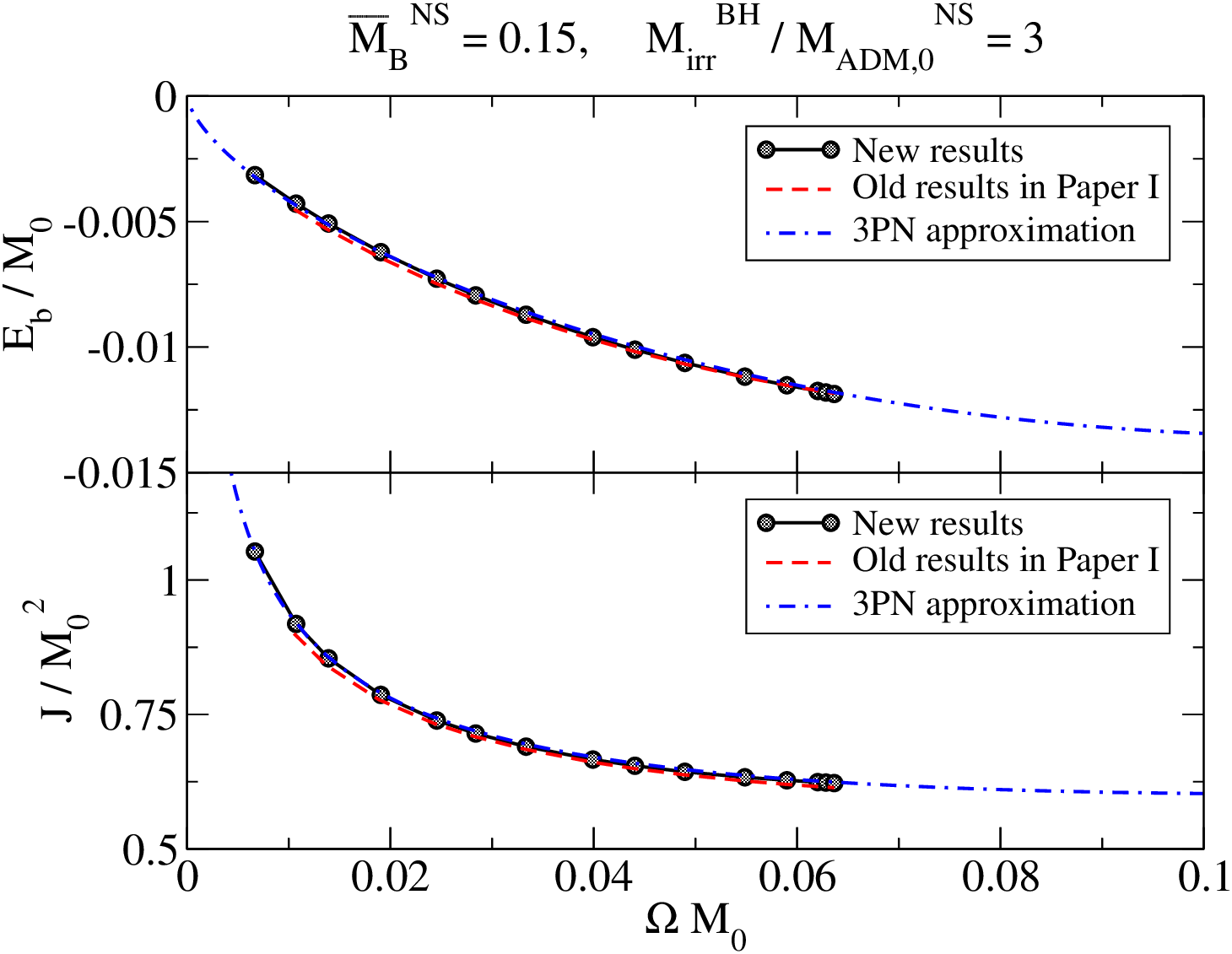}
\end{center}
\caption{Same as Fig.~\protect{\ref{fig:ej-o1M015}} but for sequences
  of mass ratio $\hat{q}=3$.}
\label{fig:ej-o3M015}
\end{figure}

\begin{figure}[ht]
\vspace{0.2cm}
\begin{center}
  \includegraphics[width=8cm]{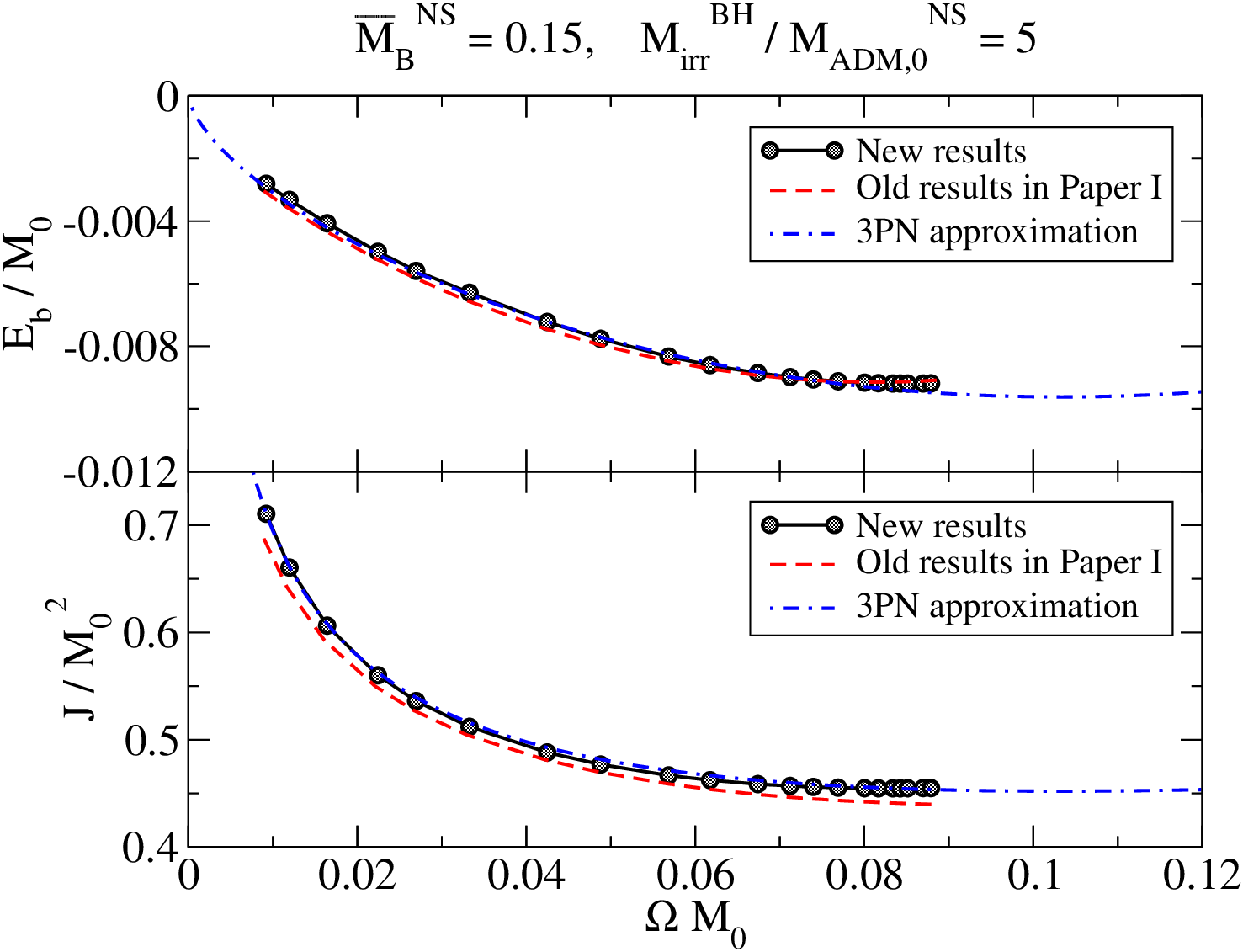}
\end{center}
\caption{Same as Fig.~\protect{\ref{fig:ej-o1M015}} but for sequences
  of mass ratio $\hat{q}=5$. The binary encounters an ISCO before the
  neutron star is tidally disrupted.}
\label{fig:ej-o5M015}
\end{figure}

\begin{figure}[ht]
\vspace{0.2cm}
\begin{center}
  \includegraphics[width=8cm]{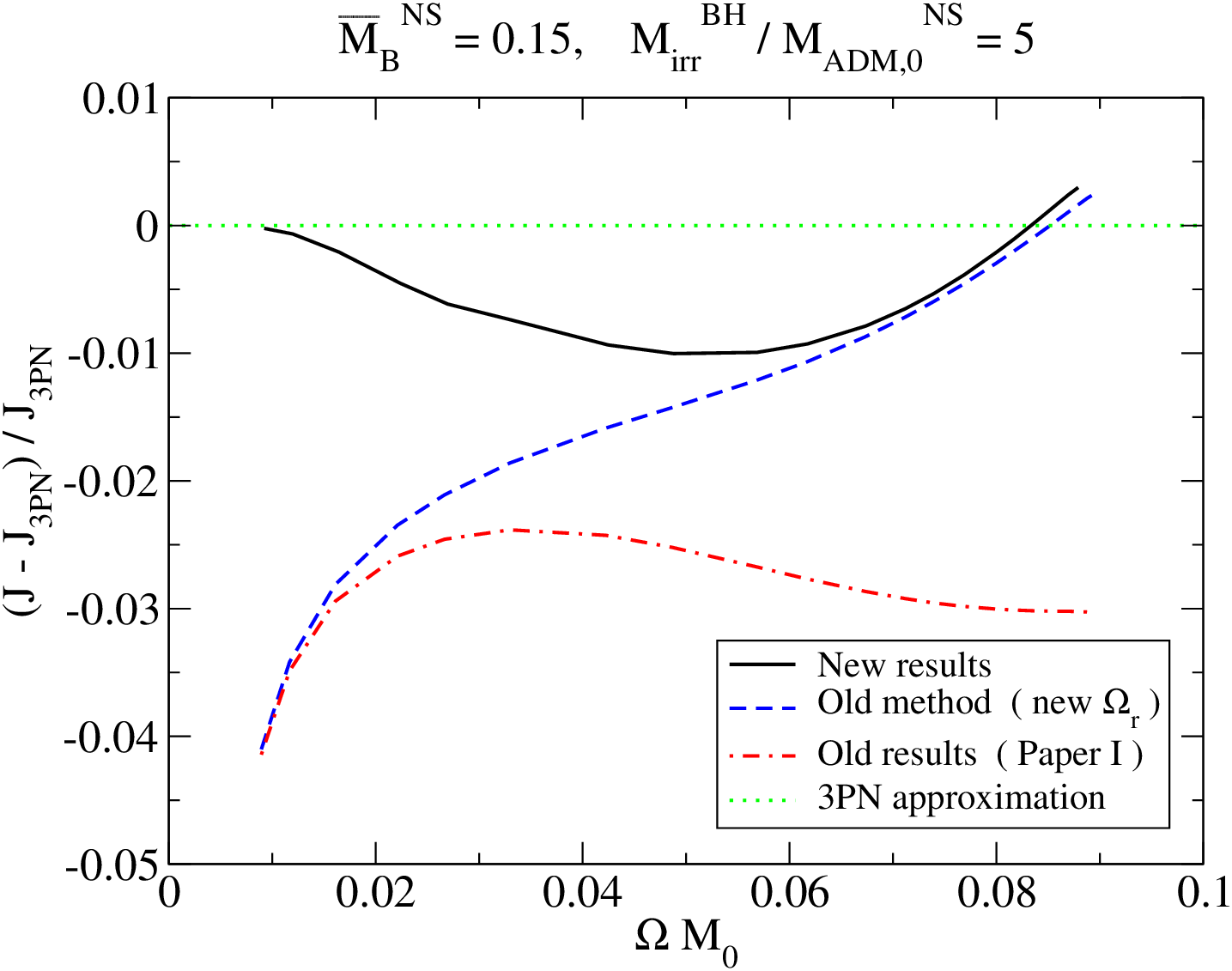}
\end{center}
\caption{The relative difference between the 3PN and numerical total
  angular momentum as a function of $\Omega M_0$ for binaries of mass
  ratio $\hat{q}=5$ and neutron-star mass
  $\bar{M}_{\rm B}^{\rm NS}=0.15$. The solid curve shows new results
  in this paper, the dashed curve the results of the old formulation
  in Paper I but using the new method to compute $\Omega_r$, and the
  dotted-dashed curve the old results in Paper I. This figure
  demonstrates that the new spin condition improves the results
  predominantly at small binary separation, while the new formulation
  of the equations has a larger effect at large binary separations.}
\label{fig:improve}
\end{figure}

In Figs.~\ref{fig:ej-o1M015} -- \ref{fig:ej-o5M015} we show the
binding energy $E_{\rm b}/M_0$ and the total angular momentum
$J/M_0^2$ as a function of the orbital angular velocity $\Omega M_0$
for neutron stars with baryon rest mass $\bar{M}_{\rm B}^{\rm
  NS}=0.15$ and for mass ratios $\hat{q} = 1$, 3, and 5. Here we
define the binding energy as the difference in the total ADM mass of
the binary from that at infinite orbital separation,
$E_{\rm b} \equiv M_{\rm ADM} - M_0$, and
the total ADM mass of the binary at infinite orbital separation as 
$M_0 \equiv M_{\rm irr}^{\rm BH} +M_{\rm ADM,0}^{\rm NS}$. In these
figures we also include, for comparison, results from Paper I and
3PN approximations \cite{Blanc02}. We find that our new results show
much better agreement with 3PN results especially for larger
separations (smaller $\Omega M_0$). This improvement is due to the
change of variables we discussed in Section \ref{Sec:IIA}.  
In Fig.~\ref{fig:improve} we show the relative difference of the total
angular momentum from that of 3PN approximation as a function of the
orbital angular velocity $\Omega M_0$ for neutron-star baryon rest
mass $\bar{M}_{\rm B}^{\rm NS}=0.15$ and mass ratio $\hat{q}=5$. The
solid curve shows the new results in this paper, the dashed curve the
results of the old formulation in Paper I but including the new method to
compute $\Omega_r$, and the dotted-dashed curve the old results in Paper
I. For closer configurations (larger $\Omega M_0$), the difference is
dominated by the accuracy of the spin parameter $\Omega_r$, while for
larger separations (smaller $\Omega M_0$), the difference comes from
the change of variables.

\begin{figure}[ht]
\vspace{0.2cm}
\begin{center}
  \includegraphics[width=8cm]{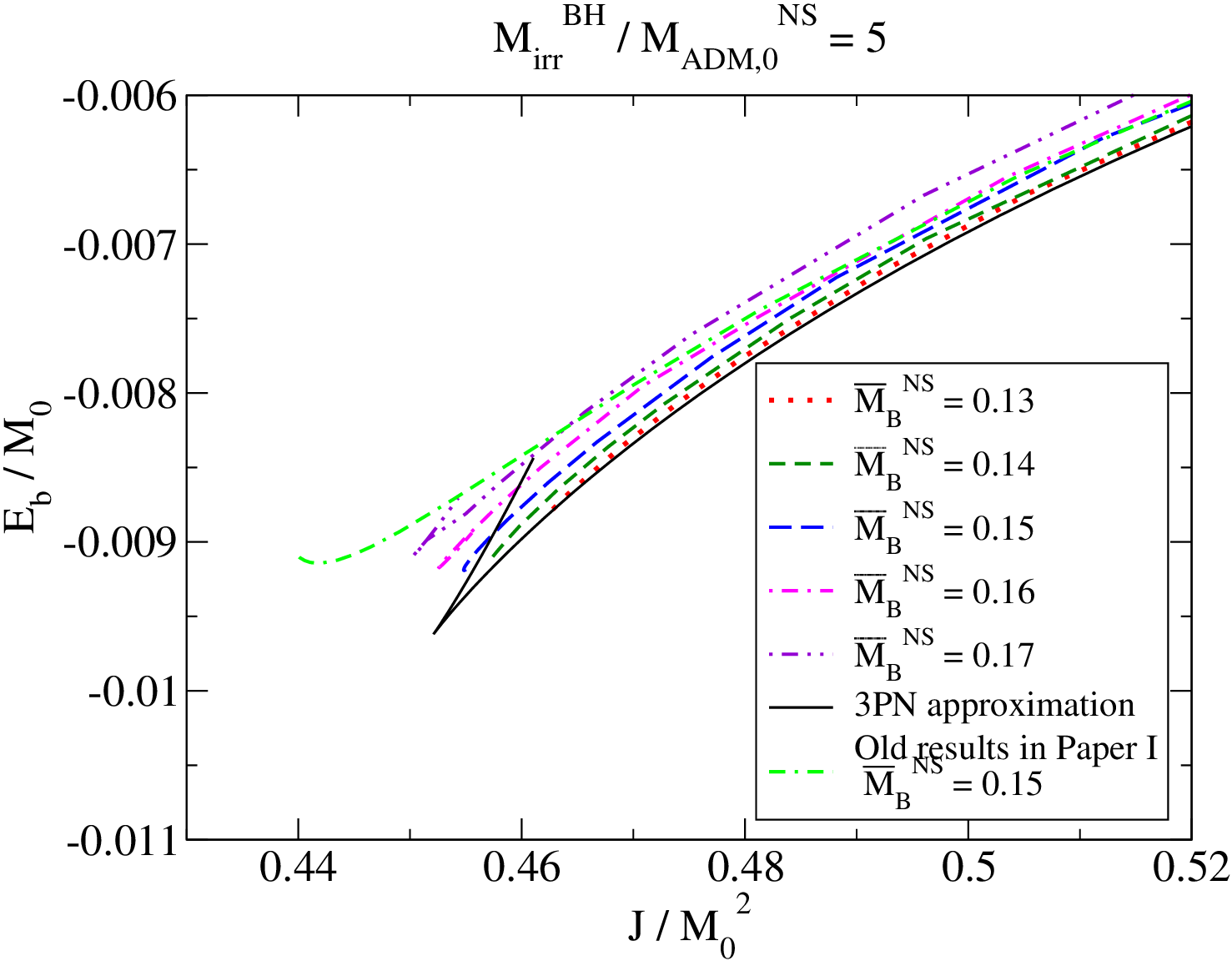}
\end{center}
\caption{The binding energy as a function of total angular momentum
  for binaries of mass ratio $\hat{q}= 5$, and different neutron star
  compactions.}
\label{fig:turning-point}
\end{figure}

Before discussing these results in greater detail it is useful to
anticipate some qualitative scaling. From a very crude Newtonian
argument, we can estimate the binary separation $d_{\rm tid}$ at which
the neutron star will be tidally disrupted by comparing the tidal
force exerted by the black hole on a test mass $m$ on the neutron
star's surface with the gravitational force exerted by the neutron
star on this test mass. Equating these two forces yields
\begin{equation} \label{eq:tidal}
  \frac{d_{\rm tid}}{M_{\rm BH}} \simeq \Bigl( \frac{M_{\rm
  NS}}{M_{\rm BH}} \Bigr)^{2/3} \frac{R_{\rm NS}}{M_{\rm NS}}.
\end{equation} 
If $d_{\rm tid}$ is greater (by a sufficient amount) than the
innermost stable circular orbit (ISCO) separation $d_{\rm ISCO}$ we
may expect the neutron star to be tidally disrupted before being
swallowed more or less intact, albeit distorted by tidal forces, by
the black hole. The qualitative relation (\ref{eq:tidal}) suggests
that the tidal separation decreases with increasing mass ratio
$\hat{q}$ and neutron star compaction.

Our sequences terminate shortly before the onset of tidal
disruption. We therefore expect to encounter minima in the binding
energy and angular momentum, which identify the ISCO, only for
binaries with sufficiently large mass ratio $\hat{q}$ and neutron star
compaction. Comparing Figs.~\ref{fig:ej-o1M015} -- \ref{fig:ej-o5M015}
we indeed find turning points only for the largest mass ratio
$\hat{q}=5$.

We also note that in this case the turning points in the binding
energy and the angular momentum occur simultaneously to within our
numerical accuracy. This was not the case with for our earlier
results, which adopted the ``leading-order" nonspinning condition
$\Omega_r = \Omega$. When imposing the more accurate condition $S = 0$
for irrotational binaries, we now do find simultaneous minima (compare
also the analogous results of \cite{CookP04,CaudiCGP06} for binary
black holes).

To highlight this finding we graph in Fig.~\ref{fig:turning-point} the
binding energy versus angular momentum for sequences of mass ratio
$\hat{q} = 5$, but different neutron star compactions. A simultaneous
turning point in the binding energy and angular momentum leads to a
cusp in these curves. As suggested by Eq.~(\ref{eq:tidal}), we do not
find turning points for small compactions (since the sequences end at
tidal disruption before encountering an ISCO), but for larger
compactions these curves indeed form a cusp. While our results agree
with 3PN results very well, we do note a small deviation that
increases with the neutron star compaction, as one might possibly
expect. We also clearly find that our new data agrees with 3PN results
much better than the old ones in Paper I for smaller binary separation
(see curves of $\bar{M}_{\rm B}^{\rm NS}=0.15$ in
Fig.~\ref{fig:turning-point}). We note that 3PN sequences cannot
identify tidal disruption and therefore always exhibit turning points.

\subsection{Quasilocal spin angular momentum of the black hole}
\label{Sec:spin}

\begin{figure}[ht]
\vspace{0.2cm}
\begin{center}
  \includegraphics[width=8cm]{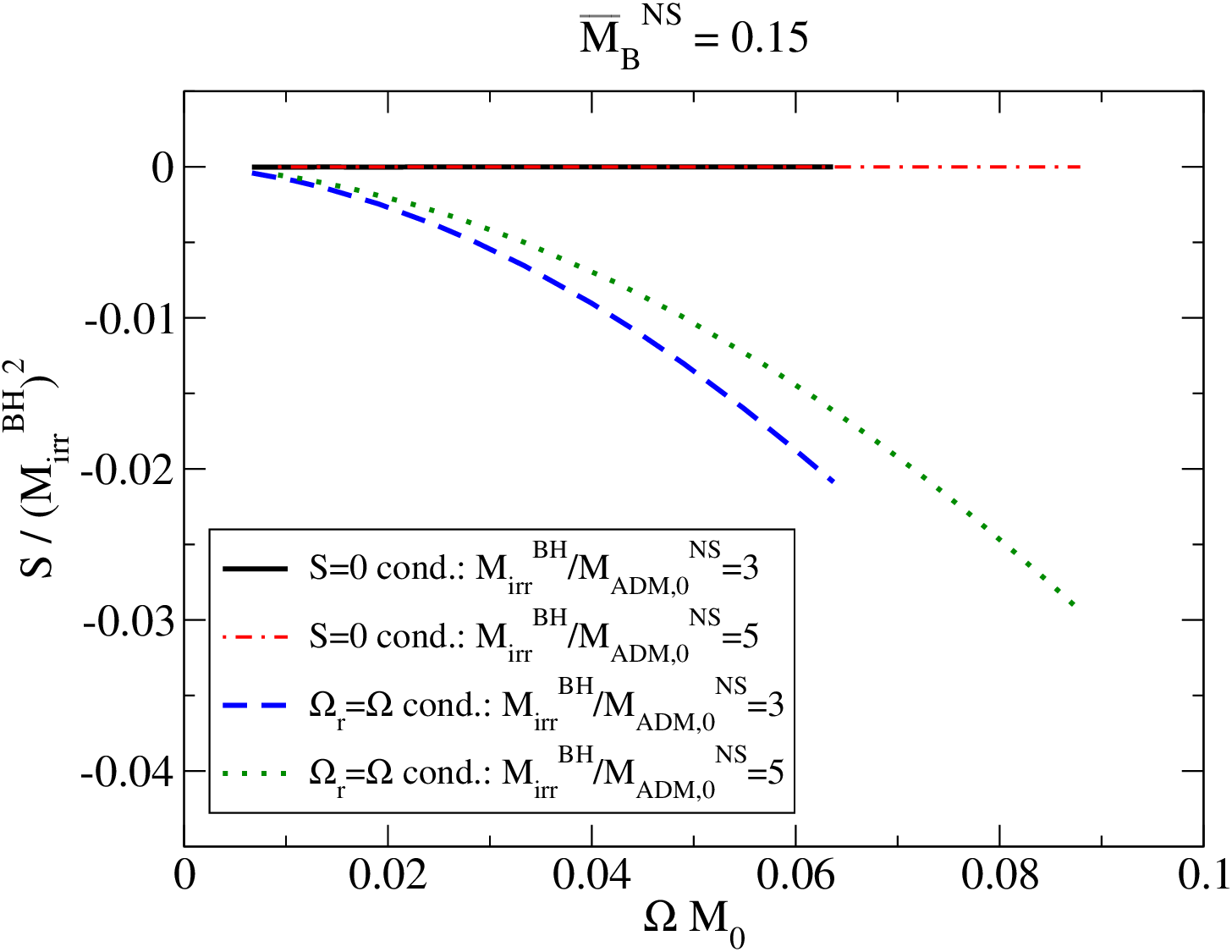}
\end{center}
\caption{Quasilocal spin angular momentum of the black hole as a
  function of the orbital angular velocity (along a quasiequilibrium
  constant-mass sequence). Solid and dotted-dashed lines correspond to
  the condition $S = 0$, while dashed and dotted curves result from
  $\Omega_r = \Omega$. (Compare the curves with the inset in Fig.~7
  in \protect{\cite{CaudiCGP06}}.)}
\label{fig:spinmom}
\end{figure}

\begin{figure}[ht]
\vspace{0.2cm}
\begin{center}
  \includegraphics[width=8cm]{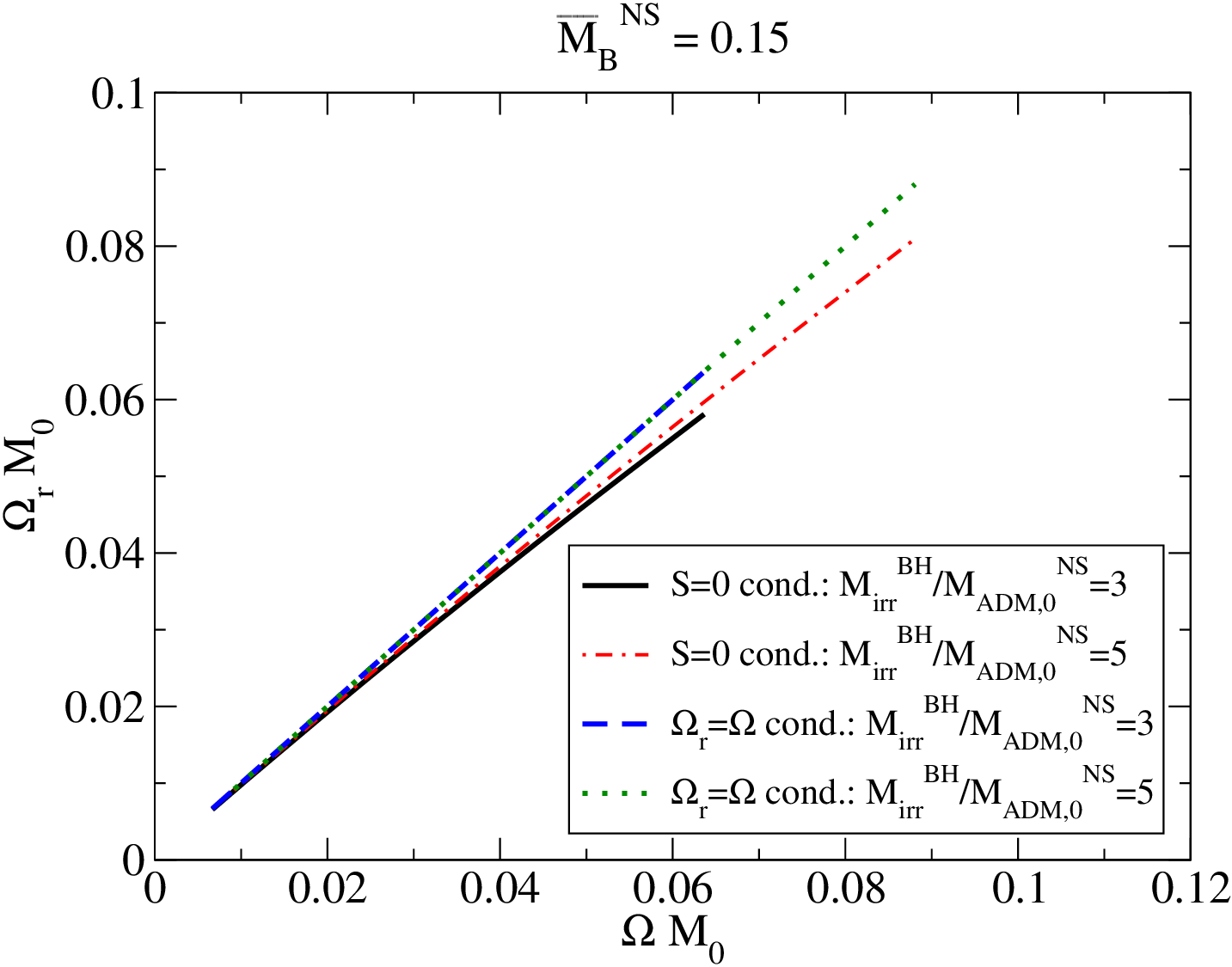}
\end{center}
\caption{Same as Fig.~\protect{\ref{fig:spinmom}}, except for the
  black hole spin parameter $\Omega_r$.}
\label{fig:spinvelo}
\end{figure}

Probably the most important numerical improvement that we present here
is the incorporation of a method to compute the spin angular velocity
of the black hole into our numerical code. We first obtain the Killing
vector on the excised surface by solving the Killing transport
equations, and then compute the quasilocal spin angular momentum
\cite{CaudiCGP06,DreyeKSS03}. Requiring the angular momentum to be
zero ($S=0$), we iterate over the black hole spin parameter
$\Omega_r$. In Paper I, by contrast, we simply set this parameter
equal to the orbital angular velocity $\Omega$, resulting in a
``leading-order" approximation. In Figs. \ref{fig:spinmom} and
\ref{fig:spinvelo}, we compare the quasilocal spin angular momentum
and black hole spin parameter $\Omega_r$, calculated under the two
conditions $\Omega_r=\Omega$ and $S=0$. For both computations, we use
our set of improved variables (as opposed to reusing our results from
Paper I). In these figures, solid and dotted-dashed lines represent
configurations with zero quasilocal spin angular momentum for the
black hole, while dashed and dotted curves represent configurations
where we set the spin angular velocity of the black hole equal to the
orbital angular velocity.

We find from Fig.~\ref{fig:spinmom} that the quasilocal spin angular
momentum of the black hole was negative and increased in magnitude as
the orbital separation decreased when we used the condition $\Omega_r
=\Omega$ in our boundary conditions. This negative spin angular
momentum results in a decrease in the total angular momentum compared
to the $S=0$ case. This explains why we did not find a minimum in the
total angular momentum along a sequence in Paper I. Here, by
maintaining $S=0$ within numerical errors, we find the minimum of the
total angular momentum is coincident with that of the binding energy.

Similarly, the fact that the quasilocal spin angular momentum becomes
negative for the condition $\Omega_r=\Omega$ implies that the spin
angular velocity $\Omega_r$ should be smaller than the orbital angular
velocity $\Omega$ when we require $S=0$, as confirmed by
Fig.~\ref{fig:spinvelo}.

\section{Qualitative Considerations}

Among the most important questions in the context of BHNS binaries is
whether the coalescence leads to a tidal disruption of the neutron
star, or whether the neutron star gets swallowed by the black hole
more or less intact, albeit distorted by tidal forces. Clearly, this
question has important consequences from the perspective of
gravitational wave observations, but perhaps even more important are
the ramifications for SGRBs. To launch such a burst requires the
formation of an accretion disk around the black hole, which can occur
only if the neutron star is disrupted prior to reaching an ISCO. To
explore this issue quantitatively requires a dynamical simulation
(compare \cite{ShibaU0607,FaberBST06,FaberBSTR06}), and part of the
motivation for the work presented in this paper is the construction of
suitable initial data for such calculations. In the meantime, however,
we may also use our quasiequilibrium models to obtain preliminary
estimates. Specifically, we will use our numerical results to
construct qualitative expressions that may be used to predict whether
a BHNS binary of arbitrary mass ratio and neutron star compaction
encounters an ISCO before being tidally disrupted or not.

\subsection{Tidal disruption}

\begin{figure}[ht]
\vspace{0.2cm}
\begin{center}
  \includegraphics[width=8cm]{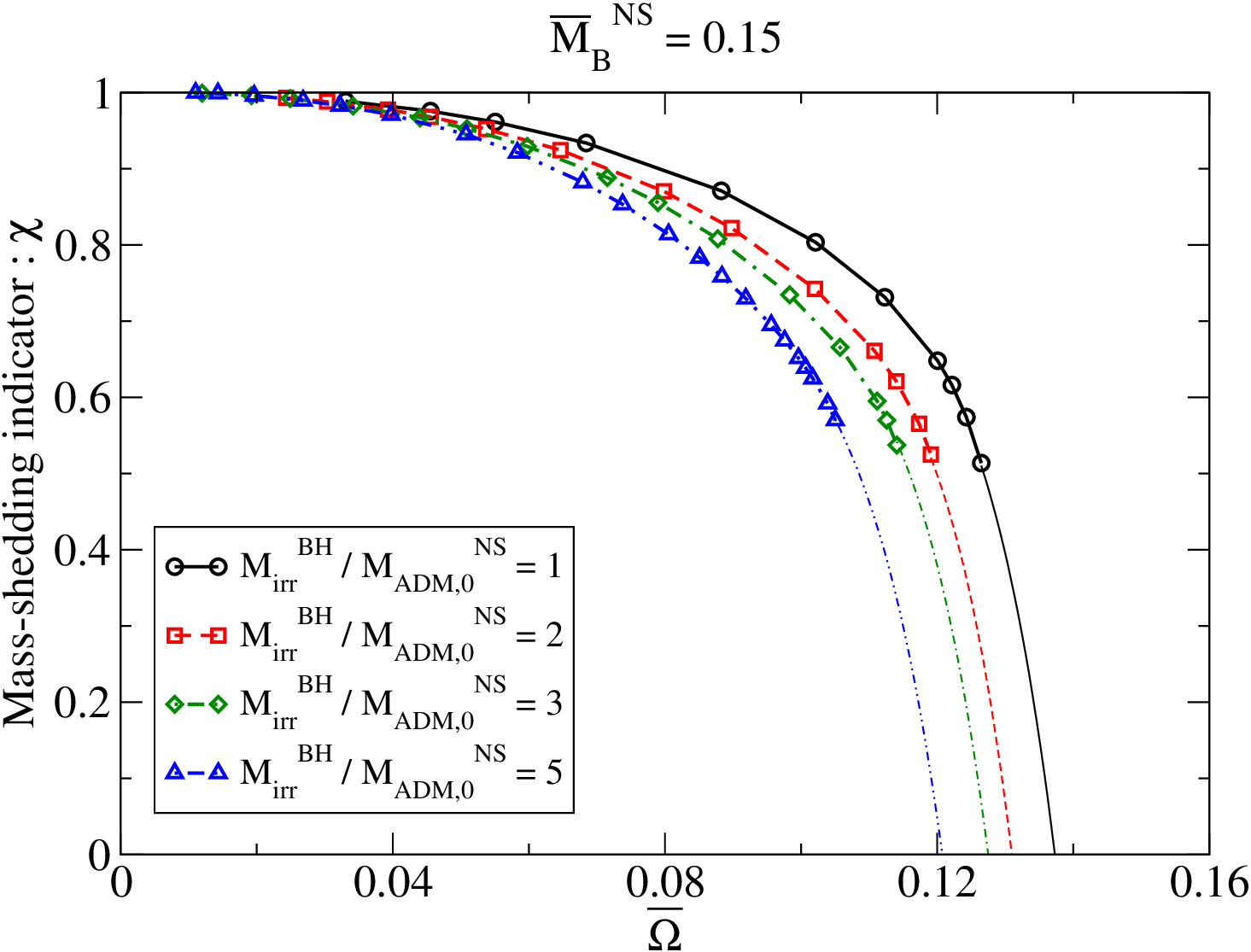}
\end{center}
\caption{Extrapolation of sequences for the neutron-star baryon rest
  mass $\bar{M}_{\rm B}=0.15$ to the mass-shedding point
  ($\chi=0$). Thick lines are sequences constructed using numerical
  data, and thin lines are extrapolated sequences.}
\label{fig:extrap}
\end{figure}

We start from investigating the binary separation (and the orbital
angular velocity) at which tidal disruption of the neutron star by the
black hole occurs. In Newtonian gravity and semi-relativistic
approaches, simple equations may be introduced to fit the effective
radius of a Roche lobe
\cite{Paczy71,Eggle83,WiggiL00,IshiiSM05}. Recently, Shibata and
Ury\=u introduced a fitting equation for binaries composed of a
nonspinning black hole and a corotating neutron star in general
relativity \cite{ShibaU0607}. In this section, we investigate our data
for a nonspinning black hole and an irrotational neutron star.

In order to obtain a fitting formula, we need to determine the orbital
angular velocity at the mass-shedding limit, i.e., the 
point that defines the onset of tidal disruption.
For this purpose we need
to extrapolate our data, because it is impossible to compute sequences
up to the tidal disruption point using a spectral methods code because
of Gibbs phenomena. To extrapolate our results, we introduce a
sensitive mass-shedding indicator, $\chi$, defined as the ratio
between radial derivatives of the specific enthalpy $h$ on the neutron
star surface in the direction of the companion and in the polar
direction,
\begin{equation}
  \chi \equiv \frac{(\partial (\ln h)/ \partial r)_{\rm eq}}
  {(\partial (\ln h)/ \partial r)_{\rm pole}}
\end{equation}
(see \cite{GourgGTMB01}). This indicator takes the value unity for a
spherical star and reaches zero at the formation of a cusp. We
tabulate $\chi$ as a function of the orbital angular velocity and
extrapolate to $\chi = 0$ by using fitting polynomial functions to
find the onset of tidal disruption. In Fig. \ref{fig:extrap}, we show
an example of such extrapolations for sequences of neutron-star baryon
rest mass $\bar{M}_{\rm B}=0.15$ with mass ratios 1, 2, 3, and 5. Note
that the horizontal axis in Fig. \ref{fig:extrap} is the orbital
angular velocity in polytropic units,
$\bar{\Omega} =\Omega R_{\rm poly}$.
Then we prepare the data of extrapolation for all models we compute.

\begin{figure}[ht]
\vspace{0.2cm}
\begin{center}
  \includegraphics[width=8cm]{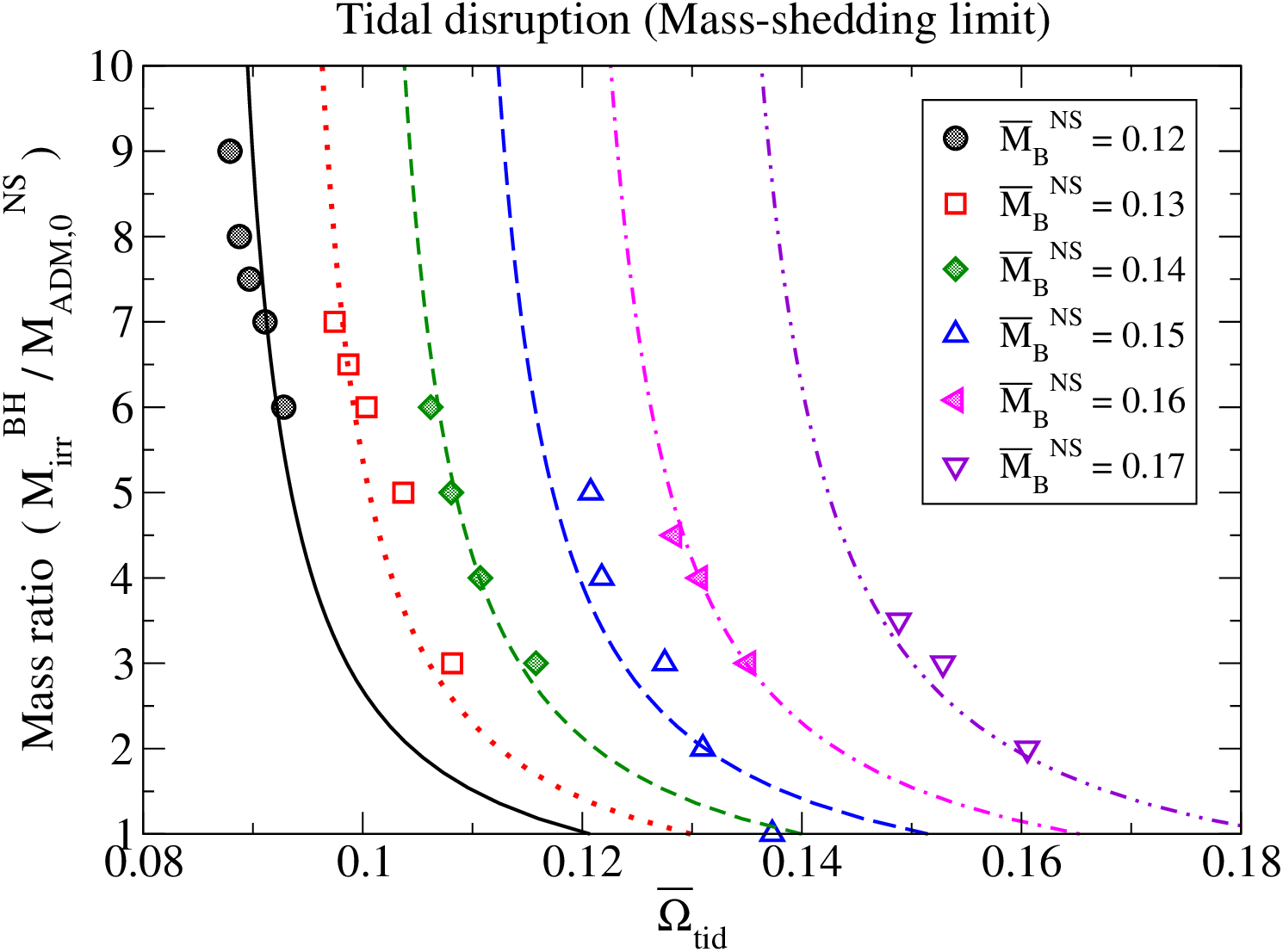}
\end{center}
\caption{Fits of the mass-shedding limit by the analytic expression
  (\protect\ref{eq:mass-shed}).
  The mass-shedding limit for each neutron-star mass and mass ratio is
  computed by the extrapolation of the numerical data.}
\label{fig:extrapall}
\end{figure}

To obtain the fitting formula for the orbital angular velocities at
the mass-shedding limit acceptable for all models we compute, we start
with the qualitative Newtonian expression (\ref{eq:tidal}), together
with Kepler's third law
\begin{equation} \label{eq:Kepler}
  \Omega \simeq \Bigl( \frac{M_{\rm BH} +M_{\rm NS}}{d_{\rm tid}^3}
  \Bigr)^{1/2}.
\end{equation}
Combining Eqs.~(\ref{eq:tidal}) with (\ref{eq:Kepler}) we can eliminate
$d_{\rm tid}$ and find
\begin{equation}
  \bar{\Omega}_{\rm tid} = 0.270
  \frac{{\cal C}^{3/2}}{\bar{M}_{\rm ADM,0}^{\rm NS}}
  \Bigl( 1 + \frac{1}{\hat{q}} \Bigr)^{1/2}, \label{eq:mass-shed}
\end{equation}
or equivalently
\begin{equation}
  \Omega_{\rm tid} M_0 = 0.270 {\cal C}^{3/2} (1+ \hat{q})
  \Bigl( 1 + \frac{1}{\hat{q}} \Bigr)^{1/2}. \label{eq:mass-shed_om}
\end{equation}
Here we have identified $M_{\rm BH}$ with $M_{\rm irr}^{\rm BH}$,
$M_{\rm NS}$ with $M_{\rm ADM,0}^{\rm NS}$, and $R_{\rm NS}$ with the
circumferential radius of a spherical neutron star $R_0$. Lastly, we
determined the coefficient of 0.270 from our numerical
results. Fitting our numerical values for the angular velocity at
tidal disruption to the form (\ref{eq:mass-shed}) resulted in a narrow
range of coefficients between 0.266 and 0.273, with a mean value of
0.270. In Fig.~\ref{fig:extrapall}, we show the results of the fitting
of the mass-shedding limit by Eq. (\ref{eq:mass-shed}).

It may be of interest to compare Eq.~(\ref{eq:mass-shed}) with a similar
expression of Shibata and Ury\=u \cite{ShibaU0607}, who express the
critical mass ratio $\hat{q}$ as
\begin{equation}
  \frac{1}{\hat q_{\rm SU}} = 0.35 \Bigl( \frac{R_{\rm NS}}{5 M_{\rm NS}}
  \Bigr)^{-3/2} \frac{(M_{\rm BH} \Omega_{\rm tid,SU})^{-1}}{6^{3/2}}
\end{equation}
(in \cite{ShibaU0607}, the mass ratio $q_{\ast}$ is defined as the
inverse of our definition). In terms of the quantities used in 
Eq.~(\ref{eq:mass-shed}), this becomes
\begin{equation}
  \bar{\Omega}_{\rm tid,SU} = 0.27
  \frac{{\cal C}^{3/2}}{\bar{M}_{\rm ADM,0}^{\rm NS}},
\end{equation}
which agrees with Eq.~(\ref{eq:mass-shed}) for large $\hat{q}$.
The agreement of our result (irrotation for the neutron star) with
that by Shibata and Ury\=u (corotation) confirms our prediction
in the limit of extreme mass ratios \cite{TanigBFS05}.

\subsection{Innermost stable circular orbit}

We are also interested in the binary separation (and the corresponding
orbital angular velocity) at which the minimum of the binding energy
appears, corresponding to the ISCO. In our numerical data we locate
this point by fitting a second-order polynomial to three nearby
points.

\begin{figure}[ht]
\vspace{0.2cm}
\begin{center}
  \includegraphics[width=8cm]{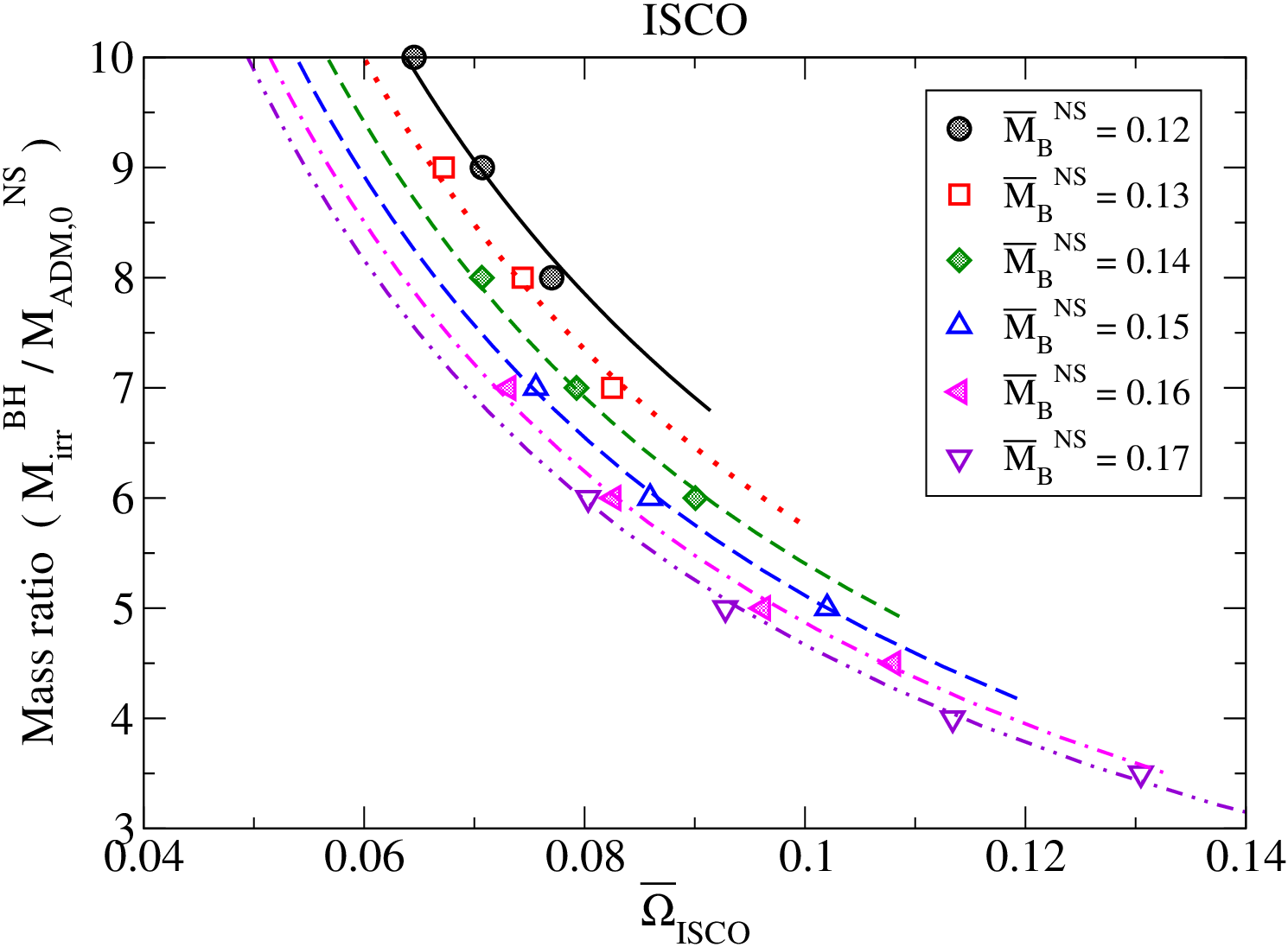}
\end{center}
\caption{Fits of the minimum point of the binding energy curve by
  expression (\protect\ref{eq:isco}).}
\label{fig:minimum}
\end{figure}

We may construct a simple empirical fit that predicts the ISCO angular
velocity $\Omega_{\rm ISCO}$ for an arbitrary companion orbiting a
black hole as follows. We search for expressions with three free
parameters that express $\Omega_{\rm ISCO}$ as a function of the mass
ratio $\hat{q}$ and the compaction ${\cal C}$ of the companion. We
then fix the three parameters by matching to three known values of
$\Omega_{\rm ISCO}$, namely (1) that of a test particle orbiting a
Schwarzschild black hole, $\Omega_{\rm ISCO} M_0=6^{-3/2}$ (for
$\hat{q} =\infty$), (2) that of an equal-mass binary black hole system
as computed in \cite{CaudiCGP06}, $\Omega_{\rm ISCO} M_0 =0.1227$ (for
$\hat{q}=1$ and ${\cal C} = 0.5$), and finally (3) that of our BHNS
configuration with $\bar{M}_{\rm B}^{\rm NS}=0.15$ and
${\cal C}=0.1452$ and mass ratio $\hat{q}=5$, resulting in 
$\Omega_{\rm ISCO} M_0 = 0.0854$. A further requirement arises from
the fact that for a test particle (with $\hat{q}=\infty$), the
expression should be independent of the companion's compaction. We
find a good fit to our remaining numerical data with the expression
\begin{equation}
  \Omega_{\rm ISCO} M_0 = 0.0680 \left( 1 - \frac{0.444}{\hat{q}^{0.25}}
  \left( 1 - 3.54 {\cal C}^{1/3} \right) \right),
  \label{eq:isco}
\end{equation}
as demonstrated in Fig.~\ref{fig:minimum}.
We determined the exponent of $\hat{q}$ and ${\cal C}$ in
Eq.~(\ref{eq:isco}) empirically by requiring that the fitted curves
lie near the data points for all models. Clearly the agreement is not
perfect, but adequate for our purposes.

\vspace{0.3cm}

\subsection{Critical mass ratio}

\begin{figure}[ht]
\vspace{0.2cm}
\begin{center}
  \includegraphics[width=8cm]{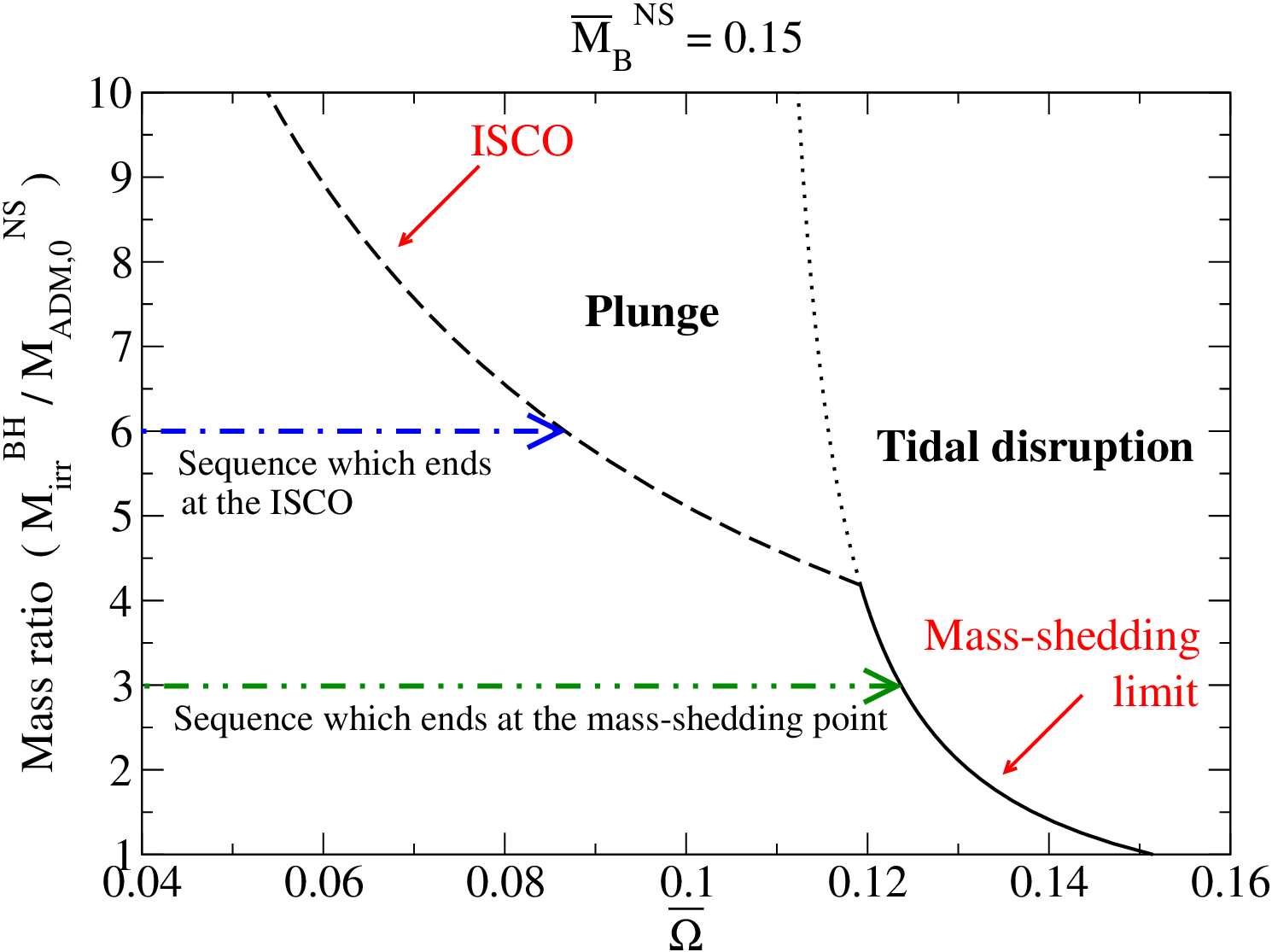}
\end{center}
\caption{An example of the boundary between the mass-shedding limit
  and the ISCO. We select the model of neutron star mass $\bar{M}_{\rm
  B}^{\rm NS}=0.15$. The solid curve denotes the mass-shedding limit,
  and the long-dashed one the ISCO for each mass ratio as a function
  of the orbital angular velocity in polytropic units. The dotted
  curve denotes the mass-shedding limit for unstable quasiequilibrium
  sequences.}
\label{fig:endpoint_example}
\end{figure}

Combining Eqs.~(\ref{eq:mass-shed}) (or (\ref{eq:mass-shed_om}))
and (\ref{eq:isco}), we can now identify the critical binary
parameters that separate those binaries that encounter an ISCO before
reaching mass-shedding, and vice-versa. In
Fig.~\ref{fig:endpoint_example} we show an example for
$\bar{M}_{\rm B}^{\rm NS}=0.15$. The solid curve denotes the orbital
angular velocity of the mass-shedding limit, and the long-dashed for
the ISCO. As seen from Eqs.~(\ref{eq:mass-shed}) and (\ref{eq:isco}),
both of these curves depend on the mass ratio, but in different ways,
which leads to the intersection of the two curves. An inspiraling
binary evolves along horizontal lines towards increasing
$\bar{\Omega}$, starting at the left and moving toward the right,
until reaching either the ISCO or the mass-shedding limit. After the
binary reaches the ISCO for sufficiently large mass ratio, we cannot
determine the exact position at which the binary disrupts via
quasiequilibrium calculations because it is in the dynamical plunge
region, but we nevertheless include the mass-shedding limit for
unstable quasiequilibrium sequences as the dotted curve in
Fig.~\ref{fig:endpoint_example}. As shown in
Fig.~\ref{fig:endpoint_example}, the model with mass ratio $\hat{q}=6$
(dotted-dashed line) encounters the ISCO, while that of $\hat{q}=3$
(dot-dot-dashed line) ends up at the mass-shedding limit. The
intersection between the mass-shedding and ISCO curve marks a critical
point that separates the two distinct outcomes of the binary inspiral.

\begin{figure}[ht]
\vspace{0.6cm}
\begin{center}
  \includegraphics[width=8cm]{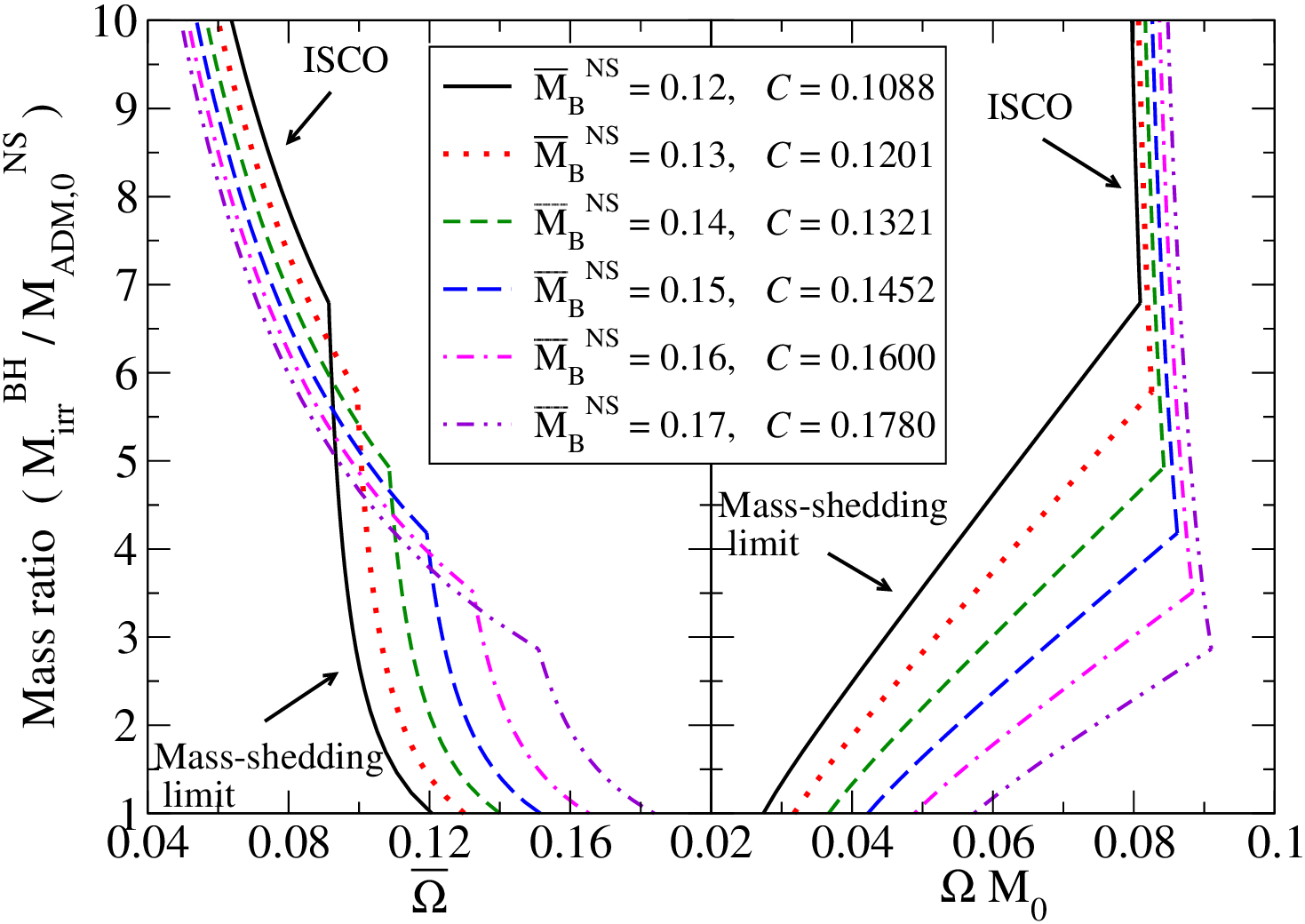}
\end{center}
\caption{Endpoint of sequences for different neutron star masses. Left
  panel is the mass ratio as a function of the orbital angular
  velocity in polytropic units, while the right panel is normalized by
  the ADM mass at infinite orbital separation.}
\label{fig:endpoint}
\end{figure}

Since expressions (\ref{eq:mass-shed}) and (\ref{eq:isco}) also depend
on the neutron star's compaction, the specific values of the critical
point likewise depend on this compaction. In Fig. \ref{fig:endpoint},
we graph the mass-shedding and ISCO curves for a number of different
neutron star masses which have a one-to-one correspondence to the
compaction. The intersections, appearing as a knee in this figure,
mark the critical point for each compaction corresponding to a binary
that encounters the ISCO at the point of tidal disruption. This
critical point also gives the maximum value of $\Omega M_0$ in
quasiequilibrium for each model. We note again that the equation of
state is fixed here to be polytropic with adiabatic index
$\Gamma=2$. This implies that the results may change, i.e., the
coefficients in Eqs.~(\ref{eq:mass-shed}) and (\ref{eq:isco}) may
change, when we change the adiabatic index or the equation of state
itself. However, we find from Fig.~\ref{fig:endpoint} that the orbital
angular velocity at the ISCO in the form of $\Omega M_0$ has a narrow
range, $0.08 \alt \Omega M_0 \alt 0.09$, for all models we
compute. This implies that the fitting formula (\ref{eq:isco}) may
hold approximately even for other equations of state.

We tabulate these critical values for the neutron star compactions
considered in this paper in Table \ref{table:crit}. Whether or not
these critical points mark the true separation between the two
outcomes will have to be verified by dynamical simulations. We
nevertheless expect that these values may provide some useful
guidance.

\begin{table}[ht]
\caption{Critical value at which tidal disruption occurs at the ISCO
  for different neutron star models.}
\begin{center}
\begin{tabular}{ccc|lccc} \hline\hline
  $\bar{M}_{\rm B}^{\rm NS}$&$\bar{M}_{\rm ADM,0}^{\rm NS}$&
  ${\cal C}$&$\bar{\Omega}_c$&$(\Omega M_0)_c$&$\hat{q}_c$&
  $(\bar{M}_{\rm irr}^{\rm BH})_c$ \\ \hline
  0.12& 0.1136& 0.1088& 0.0914& 0.0809& 6.79& 0.772 \\
  0.13& 0.1223& 0.1201& 0.0995& 0.0825& 5.78& 0.708 \\
  0.14& 0.1310& 0.1321& 0.109&  0.0843& 4.93& 0.645 \\
  0.15& 0.1395& 0.1452& 0.119&  0.0862& 4.18& 0.583 \\
  0.16& 0.1478& 0.1600& 0.132&  0.0883& 3.51& 0.519 \\
  0.17& 0.1560& 0.1780& 0.151&  0.0910& 2.86& 0.447 \\
 \hline
\end{tabular}
\end{center}
\label{table:crit}
\end{table}

\begin{figure}[ht]
\vspace{0.4cm}
\begin{center}
  \includegraphics[width=8cm]{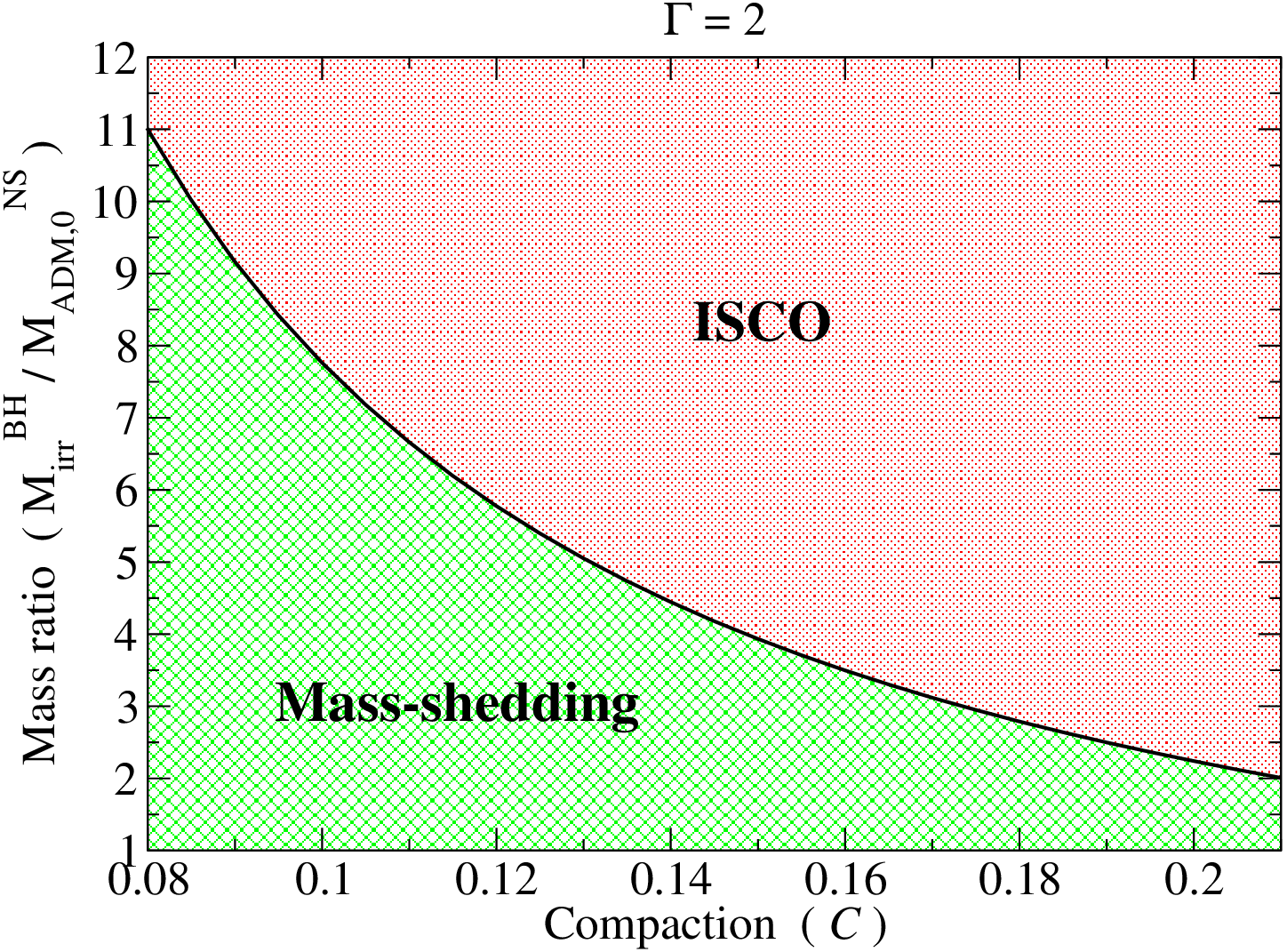}
\end{center}
\caption{Critical mass ratio which separates BHNS binaries that
  encounter an ISCO before reaching mass-shedding and undergoing tidal
  disruption, as a function of the compaction of neutron star.}
\label{fig:critpoint}
\end{figure}

When we eliminate $\Omega M_0$ from Eqs.~(\ref{eq:mass-shed_om}) and
(\ref{eq:isco}), we can draw a curve of the critical mass ratio which
separates BHNS binaries that encounter an ISCO before reaching
mass-shedding, and vice-versa, as a function of the compaction of the
neutron star. We show such a critical curve that separates those two
regions in Fig.~\ref{fig:critpoint}. The solid line denotes the
critical mass ratio for each compaction. If the mass ratio of a BHNS
binary is larger than the critical one, the quasiequilibrium sequence
terminates by encountering the ISCO, while if smaller, it ends at the
mass-shedding limit of the neutron star.

\section{Summary}

We have constructed new quasiequilibrium configurations of black
hole-neutron star binaries in general relativity. We have solved the
Einstein constraint equations in the conformal thin-sandwich formalism
coupled with the equations of relativistic hydrostationary
equilibrium. In Paper I, we set the spin angular velocity parameter of
the black hole equal to that of the orbital angular velocity in order
to produce a nonspinning black hole in the ``leading-order"
approximation \cite{CookP04}, while in this paper we compute this
parameter by requiring the quasilocal spin angular momentum of the
black hole to be zero \cite{CaudiCGP06}. We have also improved the
formulation of the gravitational field equations and obtained more
accurate results than in Paper I.

As an indication of the improvements in these calculations, a
post-Newtonian analysis predicts smaller binary eccentricities for
these new BHNS models than for those computed in Paper I
(\cite{Will07}, compare \cite{BertiIW07}). In \cite{BertiIW07}, Berti
{\it et al.} fit numerical results for the binding energy and angular
momentum of binaries in circular orbit to post-Newtonian expressions
for binaries that are not necessarily in circular orbit. Deviations
between the the two approaches then lead to non-zero eccentricities in
the post-Newtonian expressions. These eccentricities are smaller for
our new results than for those of Paper I. We also remark on another
finding of \cite{BertiIW07}, namely that for a given neutron star mass
$\bar{M}_{\rm ADM,0}^{\rm NS}$ and a given value of $\Omega M_0$, the
eccentricities in BHNS models, though small, are found to be larger
than in binary neutron star models \cite{TanigG0203}. This suggests a
larger deviation from quasiequilibrium for BHNS binaries than binary
neutron stars. But, for BHNS binaries with a mass ratio of
$\hat{q} = 5$, these parameters correspond to a larger binary
separation than for binary neutron stars with a mass ratio of
$\hat{q} = 1$ (compare Eq.~(\ref{eq:Kepler})). For similar numerical
resources, this larger binary separation leads to a larger numerical
error, which may explain the larger eccentricity found by
\cite{BertiIW07}, at least in part.

In addition to recomputing several sequences we presented in Paper I,
we have constructed sequences for a wider range of neutron star masses
and binary mass ratios, employing a $\Gamma=2$ polytropic neutron-star
equation of state throughout. We computed several constant-mass
sequences, for various mass ratios and neutron star compactions, and
searched for the appearance of a cusp at the neutron star surface --
indicating the onset of tidal disruption -- and turning points on the
binding energy and angular momentum curves -- identifying the ISCO. We
also included some qualitative fits that allow for a simple prediction
of those binary parameters separating these two different outcomes of
binary coalescence. Unlike in our earlier findings, we found
simultaneous turning points along the binding energy and angular
momentum quasiequilibrium curves.

\section*{ACKNOWLEDGMENTS}

This paper was supported in part by NSF Grants PHY-0205155,
PHY-0345151, and PHY-0650377 as well as NASA Grants NNG04GK54G and
NNX07AG96G to University of Illinois at Urbana-Champaign, and NSF
Grant PHY-0456917 to Bowdoin College.

\appendix
\section{Tables of sequences}
\label{appendix:tables}

We summarize our results in Tables \ref{table:seqM017} --
\ref{table:seqM013}. In these tables, we tabulate the coordinate
orbital separation between the center of the excised surface of the
black hole and the maximum of the neutron-star baryon rest mass $d$,
orbital angular velocity $\Omega$, spin angular velocity of the black
hole $\Omega_r$, binary binding energy $E_{\rm b}$, total angular
momentum $J$, decrease in the maximum density parameter
$\delta q_{\rm max} =(q_{\rm max} -q_{\rm max,0})/q_{\rm max,0}$ from
the spherical value $q_{\rm max,0}$, minimum of the mass-shedding
indicator $\chi_{\rm min}$, and fractional difference $\delta M$
between the ADM mass $M_{\rm ADM}$ and the Komar mass $M_{\rm Komar}$
along a sequence. Here $q_{\rm max}=(P/\rho_0)_{\rm max}$. Recall that
(virial) equilibrium requires $M_{\rm Komar}=M_{\rm ADM}$.

\begin{table*}[ht]
\caption{Physical parameters for a binary sequence with neutron-star
  baryon rest mass $\bar{M}_{\rm B}^{\rm NS}=0.17$. The ADM mass and
  the isotropic coordinate radius of the neutron star in isolation are
  $\bar{M}_{\rm ADM,0}^{\rm NS}=0.1560$ and $\bar{r}_0=0.7118$
  ($\kappa=1$). The neutron star compaction is ${\cal C}=0.1780$. We
  list the coordinate orbital separation $d$, orbital angular velocity
  $\Omega$, spin angular velocity of the black hole $\Omega_r$,
  binding energy $E_{\rm b}$, total angular momentum $J$, decrease of
  the maximum density parameter $\delta q_{\rm max}$, minimum of the
  mass-shedding indicator $\chi_{\rm min}$, and fractional difference
  $\delta M$ between the ADM mass and the Komar mass. The prefix
  $\dagger$ denotes the data closest to the position of the turning
  point, if it occurs prior to the onset of tidal disruption.}
\begin{center}
\begin{tabular}{rccllccc} \hline\hline
  \multicolumn{8}{c}{Mass ratio:
    $M_{\rm irr}^{\rm BH}/M_{\rm ADM,0}^{\rm NS}=3$} \\
  $d/M_0$&$\Omega M_0$&$\Omega_r M_0$&$E_{\rm b}/M_0$&$J/M_0^2$&
  $\delta q_{\rm max}$&$\chi_{\rm min}$&$\delta M$ \\ \hline
9.76 &0.0287 &0.0273 &-7.86(-3) &0.712 &-7.55(-3) &0.981 &1.71(-4) \\
7.81 &0.0388 &0.0364 &-9.38(-3) &0.668 &-1.22(-2) &0.958 &7.78(-5) \\
6.35 &0.0511 &0.0473 &-1.08(-2) &0.637 &-1.91(-2) &0.913 &2.22(-5) \\
5.37 &0.0634 &0.0578 &-1.19(-2) &0.619 &-2.72(-2) &0.845 &2.96(-5) \\
4.88 &0.0715 &0.0646 &-1.24(-2) &0.612 &-3.36(-2) &0.782 &4.06(-5) \\
4.59 &0.0773 &0.0693 &-1.26(-2) &0.609 &-3.94(-2) &0.724 &3.02(-5) \\
4.40 &0.0816 &0.0728 &-1.278(-2) &0.6073 &-4.46(-2) &0.669 &4.65(-6) \\
4.30 &0.0838 &0.0746 &-1.285(-2) &0.6066 &-4.79(-2) &0.626 &2.23(-5) \\
\hline
  \multicolumn{8}{c}{Mass ratio:
    $M_{\rm irr}^{\rm BH}/M_{\rm ADM,0}^{\rm NS}=5$} \\
  $d/M_0$&$\Omega M_0$&$\Omega_r M_0$&$E_{\rm b}/M_0$&$J/M_0^2$&
  $\delta q_{\rm max}$&$\chi_{\rm min}$&$\delta M$ \\ \hline
7.81 &0.0387 &0.0371 &-6.67(-3) &0.496 &-1.49(-2) &0.983 &4.33(-4) \\
5.86 &0.0564 &0.0532 &-8.12(-3) &0.466 &-2.73(-2) &0.951 &3.29(-4) \\
4.88 &0.0711 &0.0663 &-8.81(-3) &0.455 &-3.98(-2) &0.909 &3.46(-4) \\
4.30 &0.0833 &0.0768 &-9.080(-3)&0.4504&-5.61(-2) &0.856 &3.86(-4) \\
$\dagger$
4.17 &0.0865 &0.0795 &-9.096(-3)&0.4503&-6.01(-2) &0.839 &3.97(-4) \\
4.04 &0.0899 &0.0824 &-9.084(-3)&0.4505&-6.44(-2) &0.821 &4.13(-4) \\
3.71 &0.0995 &0.0904 &-8.88(-3) &0.453 &-7.70(-2) &0.761 &4.79(-4) \\
3.58 &0.1038 &0.0940 &-8.71(-3) &0.454 &-8.30(-2) &0.729 &5.12(-4) \\
\hline
\end{tabular}
\end{center}
\label{table:seqM017}
\end{table*}

\begin{table*}[ht]
\caption{Same as Table \ref{table:seqM017} but for the neutron-star
  baryon rest mass $\bar{M}_{\rm B}^{\rm NS}=0.16$. The ADM mass and
  the isotropic coordinate radius of the neutron star in isolation are
  $\bar{M}_{\rm ADM,0}^{\rm NS}=0.1478$ and $\bar{r}_0=0.7691$
  ($\kappa=1$). The neutron star compaction is ${\cal C}=0.1600$.}
\begin{center}
\begin{tabular}{rccllccc} \hline\hline
  \multicolumn{8}{c}{Mass ratio:
    $M_{\rm irr}^{\rm BH}/M_{\rm ADM,0}^{\rm NS}=3$} \\
  $d/M_0$&$\Omega M_0$&$\Omega_r M_0$&$E_{\rm b}/M_0$&$J/M_0^2$&
  $\delta q_{\rm max}$&$\chi_{\rm min}$&$\delta M$ \\ \hline
12.36 &0.0207 &0.0199 &-6.49(-3) &0.770 &-3.50(-3) &0.986 &1.09(-4) \\
 8.24 &0.0361 &0.0340 &-9.08(-3) &0.679 &-8.48(-3) &0.943 &3.46(-5) \\
 6.70 &0.0476 &0.0442 &-1.05(-2) &0.645 &-1.35(-2) &0.884 &7.31(-5) \\
 5.67 &0.0591 &0.0542 &-1.16(-2) &0.626 &-2.01(-2) &0.792 &7.37(-5) \\
 5.15 &0.0668 &0.0607 &-1.21(-2) &0.618 &-2.65(-2) &0.700 &9.46(-5) \\
 4.95 &0.0704 &0.0637 &-1.23(-2) &0.615 &-3.05(-2) &0.641 &1.21(-4) \\
 4.84 &0.0722 &0.0652 &-1.24(-2) &0.614 &-3.31(-2) &0.597 &1.52(-4) \\
 4.76 &0.0738 &0.0665 &-1.25(-2) &0.613 &-3.57(-2) &0.536 &2.19(-4) \\
 \hline
  \multicolumn{8}{c}{Mass ratio:
    $M_{\rm irr}^{\rm BH}/M_{\rm ADM,0}^{\rm NS}=5$} \\
  $d/M_0$&$\Omega M_0$&$\Omega_r M_0$&$E_{\rm b}/M_0$&$J/M_0^2$&
  $\delta q_{\rm max}$&$\chi_{\rm min}$&$\delta M$ \\ \hline
9.61 &0.0292 &0.0282 &-5.78(-3) &0.526 &-7.66(-3) &0.987 &2.69(-4) \\
6.18 &0.0526 &0.0498 &-7.95(-3) &0.471 &-2.04(-2) &0.936 &1.91(-4) \\
5.15 &0.0665 &0.0622 &-8.76(-3) &0.458 &-3.09(-2) &0.880 &1.29(-4) \\
4.47 &0.0795 &0.0736 &-9.13(-3) &0.4530 &-4.30(-2) &0.805 &1.43(-4) \\
$\dagger$
4.19 &0.0859 &0.0791 &-9.171(-3)&0.4526&-4.99(-2) &0.756 &1.64(-4) \\
3.99 &0.0914 &0.0837 &-9.13(-3) &0.4532&-5.65(-2) &0.707 &1.80(-4) \\
3.85 &0.0953 &0.0870 &-9.06(-3) &0.454 &-6.22(-2) &0.665 &1.81(-4) \\
3.71 &0.0996 &0.0905 &-8.94(-3) &0.456 &-6.91(-2) &0.611 &1.78(-4) \\
 \hline
\end{tabular}
\end{center}
\label{table:seqM016}
\end{table*}

\begin{table*}[ht]
\caption{Same as Table \ref{table:seqM017} but for the neutron-star
  baryon rest mass $\bar{M}_{\rm B}^{\rm NS}=0.15$. The ADM mass and
  the isotropic coordinate radius of the neutron star in isolation are
  $\bar{M}_{\rm ADM,0}^{\rm NS}=0.1395$ and $\bar{r}_0=0.8152$
  ($\kappa=1$). The neutron star compaction is ${\cal C}=0.1452$.}
\begin{center}
\begin{tabular}{rccllccc} \hline\hline
  \multicolumn{8}{c}{Mass ratio:
    $M_{\rm irr}^{\rm BH}/M_{\rm ADM,0}^{\rm NS}=1$} \\
  $d/M_0$&$\Omega M_0$&$\Omega_r M_0$&$E_{\rm b}/M_0$&$J/M_0^2$&
  $\delta q_{\rm max}$&$\chi_{\rm min}$&$\delta M$ \\ \hline
15.28 &0.0154 &0.0145 &-7.30(-3) &1.109 &-5.93(-4) &0.961 &1.92(-5) \\
13.10 &0.0191 &0.0179 &-8.34(-3) &1.049 &-9.17(-4) &0.934 &1.20(-5) \\
10.91 &0.0246 &0.0227 &-9.70(-3) &0.986 &-1.86(-3) &0.871 &1.27(-6) \\
 9.82 &0.0285 &0.0261 &-1.06(-2) &0.954 &-3.20(-3) &0.803 &2.90(-6) \\
 9.17 &0.0313 &0.0285 &-1.11(-2) &0.934 &-4.83(-3) &0.731 &1.19(-5) \\
 8.73 &0.0335 &0.0303 &-1.15(-2) &0.922 &-6.66(-3) &0.648 &2.77(-5) \\
 8.51 &0.0347 &0.0313 &-1.17(-2) &0.916 &-8.04(-3) &0.574 &4.80(-5) \\
 8.40 &0.0353 &0.0318 &-1.19(-2) &0.913 &-9.05(-3) &0.514 &7.56(-5) \\
 \hline
  \multicolumn{8}{c}{Mass ratio:
    $M_{\rm irr}^{\rm BH}/M_{\rm ADM,0}^{\rm NS}=2$} \\
  $d/M_0$&$\Omega M_0$&$\Omega_r M_0$&$E_{\rm b}/M_0$&$J/M_0^2$&
  $\delta q_{\rm max}$&$\chi_{\rm min}$&$\delta M$ \\ \hline
20.37 &0.0102 &0.00989&-4.99(-3) &1.104 &-7.31(-4) &0.993 &3.67(-5) \\
11.64 &0.0225 &0.0213 &-8.21(-3) &0.894 &-2.48(-3) &0.952 &5.67(-5) \\
 8.73 &0.0334 &0.0311 &-1.03(-2) &0.818 &-5.02(-3) &0.870 &6.10(-5) \\
 8.00 &0.0376 &0.0348 &-1.10(-2) &0.799 &-6.56(-3) &0.822 &5.38(-5) \\
 7.28 &0.0427 &0.0392 &-1.17(-2) &0.781 &-9.39(-3) &0.742 &5.84(-5) \\
 6.84 &0.0464 &0.0423 &-1.22(-2) &0.770 &-1.25(-2) &0.661 &8.16(-5) \\
 6.55 &0.0491 &0.0446 &-1.26(-2) &0.763 &-1.58(-2) &0.565 &1.24(-4) \\
 6.48 &0.0498 &0.0452 &-1.27(-2) &0.761 &-1.69(-2) &0.525 &1.48(-4) \\
 \hline
  \multicolumn{8}{c}{Mass ratio:
    $M_{\rm irr}^{\rm BH}/M_{\rm ADM,0}^{\rm NS}=3$} \\
  $d/M_0$&$\Omega M_0$&$\Omega_r M_0$&$E_{\rm b}/M_0$&$J/M_0^2$&
  $\delta q_{\rm max}$&$\chi_{\rm min}$&$\delta M$ \\ \hline
19.65 &0.0107 &0.0105 &-4.29(-3) &0.918 &-1.22(-3) &0.996 &1.36(-4) \\
13.10 &0.0191 &0.0184 &-6.22(-3) &0.786 &-2.70(-3) &0.982 &4.32(-5) \\
 9.82 &0.0284 &0.0270 &-7.94(-3) &0.715 &-5.11(-3) &0.953 &4.76(-5) \\
 7.64 &0.0399 &0.0375 &-9.61(-3) &0.666 &-9.14(-3) &0.888 &1.01(-4) \\
 6.55 &0.0490 &0.0455 &-1.06(-2) &0.643 &-1.35(-2) &0.808 &1.09(-4) \\
 6.00 &0.0549 &0.0506 &-1.12(-2) &0.633 &-1.76(-2) &0.734 &1.19(-4) \\
 5.46 &0.0620 &0.0567 &-1.18(-2) &0.624 &-2.53(-2) &0.595 &1.80(-4) \\
 5.35 &0.0636 &0.0581 &-1.19(-2) &0.622 &-2.78(-2) &0.537 &2.17(-4) \\
 \hline
  \multicolumn{8}{c}{Mass ratio:
    $M_{\rm irr}^{\rm BH}/M_{\rm ADM,0}^{\rm NS}=5$} \\
  $d/M_0$&$\Omega M_0$&$\Omega_r M_0$&$E_{\rm b}/M_0$&$J/M_0^2$&
  $\delta q_{\rm max}$&$\chi_{\rm min}$&$\delta M$ \\ \hline
18.19 &0.0120 &0.0118 &-3.32(-3) &0.6605 &-2.19(-3) &0.999 &2.48(-4) \\
11.64 &0.0224 &0.0218 &-4.98(-3) &0.5601 &-4.58(-3) &0.989 &1.89(-4) \\
 7.28 &0.0425 &0.0406 &-7.22(-3) &0.4882 &-1.29(-2) &0.945 &8.70(-5) \\
 5.46 &0.0618 &0.0580 &-8.60(-3) &0.4624 &-2.53(-2) &0.853 &1.55(-5) \\
 4.59 &0.0769 &0.0713 &-9.12(-3) &0.4554 &-3.92(-2) &0.730 &1.89(-5) \\
 4.37 &0.0817 &0.0755 &-9.177(-3) &0.45484 &-4.51(-2) &0.674 &2.30(-5) \\
$\dagger$
 4.22 &0.0851 &0.0784 &-9.192(-3) &0.45485 &-5.02(-2) &0.625 &3.25(-5) \\
 4.12 &0.0879 &0.0808 &-9.185(-3) &0.45500 &-5.50(-2) &0.570 &4.01(-5) \\
 \hline
\end{tabular}
\end{center}
\label{table:seqM015}
\end{table*}

\begin{table*}[ht]
\caption{Same as Table \ref{table:seqM017} but for the neutron-star
  baryon rest mass $\bar{M}_{\rm B}^{\rm NS}=0.14$. The ADM mass and
  the isotropic coordinate radius of the neutron star in isolation are
  $\bar{M}_{\rm ADM,0}^{\rm NS}=0.1310$ and $\bar{r}_0=0.8556$
  ($\kappa=1$). The neutron star compaction is ${\cal C}=0.1321$.}
\begin{center}
\begin{tabular}{rccllccc} \hline\hline
  \multicolumn{8}{c}{Mass ratio:
    $M_{\rm irr}^{\rm BH}/M_{\rm ADM,0}^{\rm NS}=3$} \\
  $d/M_0$&$\Omega M_0$&$\Omega_r M_0$&$E_{\rm b}/M_0$&$J/M_0^2$&
  $\delta q_{\rm max}$&$\chi_{\rm min}$&$\delta M$ \\ \hline
13.95 &0.0174 &0.0167 &-5.91(-3) &0.805 &-2.17(-3) &0.978 &6.26(-6) \\
10.46 &0.0260 &0.0249 &-7.56(-3) &0.729 &-4.18(-3) &0.943 &7.02(-5) \\
 8.14 &0.0367 &0.0346 &-9.18(-3) &0.678 &-7.78(-3) &0.865 &1.09(-4) \\
 6.98 &0.0451 &0.0420 &-1.02(-2) &0.653 &-1.23(-2) &0.766 &1.18(-4) \\
 6.63 &0.0482 &0.0448 &-1.05(-2) &0.646 &-1.48(-2) &0.714 &1.28(-4) \\
 6.39 &0.0505 &0.0468 &-1.07(-2) &0.642 &-1.70(-2) &0.668 &1.43(-4) \\
 6.16 &0.0530 &0.0490 &-1.10(-2) &0.638 &-2.01(-2) &0.604 &1.70(-4) \\
 5.99 &0.0551 &0.0508 &-1.12(-2) &0.635 &-2.32(-2) &0.530 &2.06(-4) \\
  \hline
  \multicolumn{8}{c}{Mass ratio:
    $M_{\rm irr}^{\rm BH}/M_{\rm ADM,0}^{\rm NS}=5$} \\
  $d/M_0$&$\Omega M_0$&$\Omega_r M_0$&$E_{\rm b}/M_0$&$J/M_0^2$&
  $\delta q_{\rm max}$&$\chi_{\rm min}$&$\delta M$ \\ \hline
10.85 &0.0247 &0.0240 &-5.36(-3) &0.547 &-4.95(-3) &0.979 &6.41(-5) \\
 7.75 &0.0391 &0.0374 &-6.96(-3) &0.496 &-1.08(-2) &0.934 &3.33(-6) \\
 6.20 &0.0524 &0.0496 &-8.08(-3) &0.473 &-1.85(-2) &0.859 &8.52(-5) \\
 5.43 &0.0622 &0.0584 &-8.64(-3) &0.463 &-2.61(-2) &0.777 &1.12(-4) \\
 5.04 &0.0683 &0.0639 &-8.89(-3) &0.460 &-3.23(-2) &0.707 &1.25(-4) \\
 4.73 &0.0740 &0.0688 &-9.06(-3) &0.458 &-4.00(-2) &0.619 &1.47(-4) \\
 4.58 &0.0771 &0.0716 &-9.12(-3) &0.4572&-4.54(-2) &0.546 &1.65(-4) \\
 4.54 &0.0780 &0.0723 &-9.14(-3) &0.4570&-4.72(-2) &0.516 &1.77(-4) \\
  \hline
\end{tabular}
\end{center}
\label{table:seqM014}
\end{table*}

\begin{table*}[ht]
\caption{Same as Table \ref{table:seqM017} but for the neutron-star
  baryon rest mass $\bar{M}_{\rm B}^{\rm NS}=0.13$. The ADM mass and
  the isotropic coordinate radius of the neutron star in isolation are
  $\bar{M}_{\rm ADM,0}^{\rm NS}=0.1223$ and $\bar{r}_0=0.8923$
  ($\kappa=1$). The neutron star compaction is ${\cal C}=0.1201$.}
\begin{center}
\begin{tabular}{rccllccc} \hline\hline
  \multicolumn{8}{c}{Mass ratio:
    $M_{\rm irr}^{\rm BH}/M_{\rm ADM,0}^{\rm NS}=3$} \\
  $d/M_0$&$\Omega M_0$&$\Omega_r M_0$&$E_{\rm b}/M_0$&$J/M_0^2$&
  $\delta q_{\rm max}$&$\chi_{\rm min}$&$\delta M$ \\ \hline
14.93 &0.0158 &0.0154 &-5.58(-3) &0.825 &-1.79(-3) &0.974 &1.39(-5) \\
11.20 &0.0237 &0.0227 &-7.15(-3) &0.747 &-3.51(-3) &0.933 &7.52(-5) \\
 8.71 &0.0334 &0.0316 &-8.71(-3) &0.691 &-6.94(-3) &0.841 &1.03(-4) \\
 8.09 &0.0370 &0.0348 &-9.19(-3) &0.678 &-8.82(-3) &0.792 &1.05(-4) \\
 7.47 &0.0412 &0.0386 &-9.71(-3) &0.665 &-1.19(-2) &0.717 &1.16(-4) \\
 6.97 &0.0451 &0.0421 &-1.02(-2) &0.655 &-1.63(-2) &0.617 &1.45(-4) \\
 6.85 &0.0462 &0.0431 &-1.03(-2) &0.652 &-1.79(-2) &0.580 &1.60(-4) \\
 6.72 &0.0473 &0.0441 &-1.04(-2) &0.650 &-1.97(-2) &0.531 &1.79(-4) \\
  \hline
  \multicolumn{8}{c}{Mass ratio:
    $M_{\rm irr}^{\rm BH}/M_{\rm ADM,0}^{\rm NS}=5$} \\
  $d/M_0$&$\Omega M_0$&$\Omega_r M_0$&$E_{\rm b}/M_0$&$J/M_0^2$&
  $\delta q_{\rm max}$&$\chi_{\rm min}$&$\delta M$ \\ \hline
11.62 &0.0225 &0.0219 &-5.06(-3) &0.560 &-4.13(-3) &0.975 &6.73(-5) \\
 8.30 &0.0356 &0.0342 &-6.66(-3) &0.505 &-9.10(-3) &0.923 &5.36(-5) \\
 7.47 &0.0410 &0.0392 &-7.19(-3) &0.492 &-1.18(-2) &0.890 &9.24(-5) \\
 6.64 &0.0479 &0.0456 &-7.76(-3) &0.480 &-1.61(-2) &0.835 &1.26(-4) \\
 5.81 &0.0570 &0.0538 &-8.36(-3) &0.469 &-2.36(-2) &0.735 &1.60(-4) \\
 5.40 &0.0627 &0.0589 &-8.65(-3) &0.465 &-3.06(-2) &0.645 &1.90(-4) \\
 5.23 &0.0652 &0.0611 &-8.76(-3) &0.463 &-3.45(-2) &0.590 &2.09(-4) \\
 5.11 &0.0672 &0.0629 &-8.84(-3) &0.462 &-3.81(-2) &0.532 &2.24(-4) \\
   \hline
\end{tabular}
\end{center}
\label{table:seqM013}
\end{table*}



\begin{thebibliography}{99}
\bibitem{dMOmegadJ}
An explicit proof in general relativity that simultaneous turning
points exist along quasiequilibrium sequences has been given for
corotating black holes in compact binaries \cite{FriedUS02}. No proof
has yet been provided for nonspinning black holes in binaries, but the
same result is expected to hold; see also \cite{CaudiCGP06}.
\end{thebibliography}
\end{document}